\documentclass[prb,twocolumn,superscriptaddress,longbibliography]{revtex4-2}
\usepackage{graphicx}
\usepackage{pstricks}
\usepackage{pst-node}
\usepackage{pstricks-add}
\usepackage{amsfonts,amssymb,amsmath,xspace,dsfont}
\usepackage{graphicx}
\usepackage[colorlinks=true,citecolor=green,linkcolor=red,urlcolor=blue]{hyperref}
\usepackage{subfigure}
\usepackage{calligra}
\usepackage{dsfont}
\usepackage{color}
\usepackage{ulem}
\usepackage{comment}
\usepackage[mathscr]{eucal}
\definecolor{g}{rgb}{.1,0.4,.1} 
\definecolor{b}{rgb}{0,0.2,1}
\definecolor{rouge}{rgb}{0.82,0.,0.}
\definecolor{vert}{rgb}{0.,0.82,0.}
\definecolor{orange}{rgb}{1,0.5,0.}
\definecolor{bleu}{rgb}{0.,0.,0.82}
\definecolor{m}{rgb}{0.82,0.,0.82}
\definecolor{vert2}{rgb}{0.,0.5,0.}
\definecolor{rougeclair}{rgb}{1.0,0.7,0.7}

\newcommand{\be}{\begin{equation}}
\newcommand{\ee}{\end{equation}}
\newcommand{\beqn}{\begin{eqnarray}}
\newcommand{\eeqn}{\end{eqnarray}}

\newcommand{\bem}{\begin{pmatrix}}
\newcommand{\eem}{\end{pmatrix}}

\newlength{\ldag}
\newcommand{\bra}[1]{\langle#1|}
\newcommand{\ket}[1]{|#1\rangle}

\newcommand{\tr}{\mathrm{Tr}}

\newcommand{\newZ}{{\mathscr{Z}}}

\begin{document}

\title{Finite-temperature properties of string-net models}

\author{Anna Ritz-Zwilling}
\email{anna.ritz\_zwilling@sorbonne-universite.fr }
\affiliation{Sorbonne Universit\'e, CNRS, Laboratoire de Physique Th\'eorique de la Mati\`ere Condens\'ee, LPTMC, F-75005 Paris, France}

\author{Jean-No\"el Fuchs}
\email{jean-noel.fuchs@sorbonne-universite.fr}
\affiliation{Sorbonne Universit\'e, CNRS, Laboratoire de Physique Th\'eorique de la Mati\`ere Condens\'ee, LPTMC, F-75005 Paris, France}
\
\author{Steven H. Simon}
\email{steven.simon@physics.ox.ac.uk}
\affiliation{Rudolf Peierls Centre for Theoretical Physics, Oxford, OX1 3PU, United Kingdom}

\author{Julien Vidal}
\email{julien.vidal@sorbonne-universite.fr}
\affiliation{Sorbonne Universit\'e, CNRS, Laboratoire de Physique Th\'eorique de la Mati\`ere Condens\'ee, LPTMC, F-75005 Paris, France}

\begin{abstract}
We consider a refined version of the string-net model which assigns a different energy cost to each plaquette excitation. Using recent exact calculations of the energy-level degeneracies we compute the partition function of this model and investigate several thermodynamical quantities. In the thermodynamic limit, we show that the partition function is dominated by the contribution of special particles, dubbed pure fluxons, which trivially braid  with all other (product of) fluxons. We also analyze the behavior of Wegner-Wilson loops associated to excitations and show that they obey an area law, indicating confinement, for any finite temperature except for pure fluxons that always remain deconfined. Finally, using a recently proposed conjecture, we compute the topological mutual information at finite temperature, which features a nontrivial scaling between system size and temperature, similar to the one-dimensional classical Ising model.
\end{abstract}

\maketitle


\section{Introduction}

Topological quantum phases of matter are characterized by various properties that are robust against small perturbations (see Ref.~\cite{Wen17} for a review). In two dimensions, the most famous examples are the chiral phases associated with the fractional quantum Hall effect~\cite{Tsui82}. One of the main features of these topologically ordered phases is the nontrivial quantum statistics of quasiparticles, called anyons~\cite{Leinaas77,Goldin80,Goldin81,Wilczek82_1,Wilczek82_2}, which have recently been studied in several experiments~\cite{Bartolomei20,Nakamura20,Werkmeister24,Samuelson24}.  The robustness of these phases and the exotic braiding properties of anyons make them very good candidates for quantum memories and topological quantum computing (see Refs.~\cite{Nayak08} and references therein).

To analyze topologically ordered phases, it is often useful to consider exactly solvable models such as the toric code model~\cite{Kitaev03} or the string-net (SN) model~\cite{Levin05}, to name but a few. Interestingly, these models can now be simulated using different quantum artificial devices. For instance, the toric code model has been implemented using superconducting qubits~\cite{Song18,Andersen20,Chen21,Satzinger21,Xu23}, or nuclear magnetic resonance simulator~\cite{Luo18}, whereas the SN model has very recently been designed via a superconducting quantum processor~\cite{Li17,Minev24,Xu24}. The aforementioned models are considered as the paradigmatic models of topological phases. The corresponding Hamiltonians are made of local commuting projectors and, as such, they only generate a particular class of topological phases known as achiral (vanishing Hall conductance) and doubled~\cite{Kapustin20,Zhang22}. 

Although topological quantum order is stable against local perturbations~\cite{Bravyi10} at zero temperature ($T=0$), it may be very fragile in the presence of thermal fluctuations ($T\neq 0$). In two dimensions, it has even been shown that, for any Hamiltonian built from local commuting projectors, topological order is destroyed in the thermodynamic limit for any $T>0$~\cite{Hastings11}. Nevertheless, a subtle interplay between the system size and the temperature allow to consider a size-dependent temperature below which the topological order is preserved (see, e.g., Refs.~\cite{Castelnovo07_2,Nussinov08} for a discussion of the toric code model). This problem of thermal fragility is especially important in the perspective of using these systems to realize self-correcting memories~\cite{Dennis02,Bacon06,Alicki09,Landon13,Brown16}. 

In this paper, we study a refined version of the SN model and study its finite-temperature properties. In recent years, several generalizations of the SN model have been proposed~\cite{Lan14,Lin14,Hahn20,Lin21}. These generalizations allow one to consider any input unitary fusion category (UFC) $\mathcal{C}$ and, hence, to generate all possible achiral phases. These emergent achiral topological phases are described by the Drinfeld center of $\mathcal{C}$, $\mathcal{Z(C)}$, which is a unitary modular tensor category, i.e., a well-behaved anyon theory. 
Akin to quantum double models~\cite{Kitaev03}, SN models have two types of excitations corresponding to violations of either vertex constraints or plaquette constraints. Nevertheless, by construction, SN models cannot create pure vertex excitations. In other words, when a vertex is excited, neighboring plaquettes are also automatically excited [this problem can be circumvented by adding extra degrees of freedom (tails) giving rise to the so-called extended SN models~\cite{Hu18}]. 
Here, as originally proposed by Levin and Wen~\cite{Levin05}, we restrict to the Hilbert space satisfying the vertex constraints (the branching rules) so that only plaquette excitations are allowed (see Sec.~\ref{sec:model}), but we propose a refinement of the SN model by assigning a different energy cost to each possible plaquette excitation (a similar extension has been already discussed for quantum double models in Ref.~\cite{Komar17}). Using our recent exact calculations of the energy-level degeneracies~\cite{Ritz24}, we compute various quantities at finite temperature in this refined string-net (RSN) model which encompasses, as special cases, the original SN model and its generalizations~\cite{Lan14,Lin14,Hahn20,Lin21}. 
Our main results can be summarized as follows: (i) we compute the partition function, the energy, specific heat, and  the entropy (see also Ref.~\cite{Vidal22} for related works), (ii) we obtain simple expressions of  the Wegner-Wilson loops (for contractible and noncontractible contours) at finite temperature in terms of effective partition functions, (iii) we discuss the topological mutual information and find a nontrivial scaling behavior (involving the system size and the temperature) similar to the one found in the one-dimensional classical Ising model. 
Importantly, we show that, while the Drinfeld center $\mathcal{Z}(C)$ is enough to describe the ground state of the system, additional information is needed at finite temperature, namely, the internal multiplicities of the anyons, which depend on the input category $\mathcal{C}$ and are obtained from the tube algebra. Among simple objects of $\mathcal{Z}(C)$, we identify subsets that play special roles and depend on $\mathcal{C}$: the fluxons $\mathcal{F}$ (i.e., single-plaquette excitations), a particular subset of the  fluxons, which we call {\it pure fluxons} $\mathcal{P}$, and the fusion product of fluxons, notated $\mathcal{F}^\otimes$. 

The paper is organized as follows. We first review the construction of SN models (Sec.~\ref{sec:model}) and introduce a generalization (RSN) that lifts all nontopological degeneracies in the energy spectrum. Then, we review the results on energy-level degeneracies that were obtained in~\cite{Ritz24} and add the case of a punctured manifold (Sec.~\ref{sec:deg}). Next, we compute the partition function (Sec.~\ref{sec:partfunc}) and analyze the resulting equilibrium thermodynamics (Sec.~\ref{sec:thermo}). In Sec.~\ref{sec:proj}, we compute the thermal average of projectors onto given quasiparticle sectors in a given region. Using these results for projectors, it is straightforward to compute the thermal average of Wegner-Wilson loops and to discuss the confinement of quasiparticles (Sec.~\ref{sec:wwlconf}). Then, we study entanglement properties by focusing on the topological mutual information at finite temperature (Sec.~\ref{sec:tmi}). In Sec.~\ref{sec:ccl}, we conclude and give perspectives. 
Appendices provide details on the notions of fluxons (Appendix~\ref{app:transpflu}), on the fusion of simple objects of the Drinfeld center as obtained from the tube algebra (Appendix~\ref{app:fusiontubes}), and on a surgery approach to computing the degeneracies (Appendix~\ref{app:surg}).

%
%
\section{String-net models}
\label{sec:model}
%

The original SN model has been introduced by Levin and Wen~\cite{Levin05} and generalized in Refs.~\cite{Lan14,Lin14,Hahn20,Lin21} in order to consider any UFC as input category. In the following, we briefly recall the basic ingredients of this model and we refer the interested reader to Ref.~\cite{Lin21} for a more complete and detailed introduction (see also Ref.~\cite{Ritz24} for a brief overview). Below, we shall propose another extension of the SN models, which can also be defined for any UFC. In some sense, it is a ``refined generalized" SN model but for simplicity, we will just call it RSN model. 

A SN model is built from a UFC $\mathcal{C}$ whose fusion rules define the Hilbert space of the system. It is defined on a two-dimensional trivalent graph. Here, for simplicity, we consider a trivalent graph with $N_{\rm  p}$ plaquettes embedded on an orientable closed surface. The case of surfaces with boundaries is discussed in Ref.~\cite{Ritz24} and subtleties arising for nonorientable surfaces are addressed, e.g.,  in Refs.~\cite{Chan16,Barkeshli20}.

A basis for the Hilbert space is obtained by assigning a quantum number $a,b,c,\ldots$ to each directed edge of the graph. These microscopic degrees of freedom are chosen as simple objects of the category  $\mathcal{C}$. In some cases, extra quantum numbers may also be assigned to vertices.  The UFC is equipped with fusion multiplicity coefficients $N_{abc}$ (sometimes written in the form of a matrix $N_a$ with rows $b$ and columns $c$) which give the number of ways that $a,b,c$ can fuse to the identity. That is, if $a,b,c$ meet at a vertex (inwardly directed) and fuse to the identity, then the vertex is given a quantum number which  ranges from $1$ to  $N_{abc}$. If $a,b,c$ do not fuse to the identity, then $N_{abc}=0$.  

In this work, we restrict the Hilbert space to basis states satisfying the fusion rules at each trivalent vertex. In other words, we exclude vertex configurations with $N_{abc}=0$.  

%
\subsection{String-net model}
\label{subsec:SN}
%
Within the restricted Hilbert space $\mathcal{H}$, the original Hamiltonian of the SN model is given by~\cite{Levin05}: 
%
%
\be
H = - \sum_{p=1}^{N_{\rm  p}} B_p, 
\label{eq:ham}
\ee 
%
%
where $B_p$'s are local commuting projectors acting on the plaquette $p$. The ground-state manifold is generated by all states $|\psi \rangle$ such that $B_p|\psi \rangle=|\psi \rangle$ for all plaquettes $p$. Excited states are obtained by violating this plaquette constraint. Elementary excitations can thus be seen as plaquette excitations and are dubbed fluxons~\cite{Ritz24}. For a given input UFC $\mathcal{C}$,  these fluxons belong to a subset $\cal F$ of the simple objects of the so-called Drinfeld center $\mathcal{Z(C)}$ (by convention, we also include the vacuum ${\bf 1}$ in this subset although it does not cost any energy).  Hence, $B_p$ can be interpreted as the projector onto the vacuum of $\mathcal{Z(C)}$, inside the plaquette $p$. The excitation energy of a plaquette containing a fluxon $A \neq \bf 1$ is set to 1. Note that, because of the vertex constraint, some objects of $\mathcal{Z(C)}$ may not be generated even by multiple fusion of fluxons (see Sec.~\ref{sub:cut}). 

%
\subsection{Refined string-net model}
\label{subsec:RSN}
%
Actually, one can consider a refined version of the SN model by assigning to each plaquette $p$, a Hermitian coupling matrix $\mathcal{J}_p^{A}$, which depends on the fluxon type $A$ and whose row and column indices $a,b$ correspond to internal multiplicity indices. The notion of an internal multiplicity is only relevant for noncommutative UFCs where a fluxon $A$ may carry a multiplicity index $a$ running from 1 to $n_{A,1}>1$~\cite{Ritz24}. In the commutative case, one always has $n_{A,1}=1$ for fluxons (and $0$ otherwise). 

The RSN Hamiltonian is then given by:
%
%
\be
H=- \sum_{p=1}^{N_{\rm p}}  \sum_{A \in {\cal F}}  \sum_{a,b=1}^{n_{A,1}} \mathcal{J}_p^{A,ab} B_{p}^{A,ab},
\label{eq:RSN}  
\ee
%
%
where $B_{p}^{A,ab}$ acts in the $n_{A,1}$-dimensional internal space of fluxon $A$ in the plaquette $p$. When $a=b$, this operator corresponds to the simple idempotent (projector) $p_A^{11,aa}$ of the tube algebra in the $11$ sector, acting in the plaquette $p$. Similarly, when  $a\neq b$, $B_{p}^{A,ab}$ corresponds to the nilpotent $p_A^{11,ab}$ in the $11$ sector (see Ref.~\cite{Ritz24} for more details), acting in the plaquette $p$. The matrix elements of the operators $B_p^{A,ab}$ in the edge basis can be computed using the tube algebra (see, e.g., Refs.~\cite{Lan14,Ritz24}).
The fact that we only consider operators acting in the $11$ sector is simply due to the vertex constraints (Gauss law) which are always satisfied in the restricted Hilbert space.

For the  vacuum $A={\bf 1}$, which has no internal multiplicity ($n_{\mathbf{1},1}=1$), the corresponding projector is the same as the one introduced in Eq.~\eqref{eq:ham} ($B^{\mathbf{1},11}_p=B_p$). The SN model is then recovered by choosing $\mathcal{J}_p^{A,ab}=\delta_{A,\mathbf{1}}$, for all  $p$'s. We stress that other choices of $\mathcal{J}_p^{A,ab}$ may lead to ground states with nontrivial fluxons. 

In the following, only the eigenvalues of the matrix $\mathcal{J}_p^{A}$ will appear and will be called $J_p^{A,a}$ where $a$ runs from 1 to $n_{A,1}$. In the basis that diagonalizes the coupling matrix, the Hamiltonian~(\ref{eq:RSN}) can be rewritten as a sum of local commuting projectors (see also Appendix D of Ref.~\cite{Ritz24}) and therefore remains exactly solvable. A key feature of the RSN Hamiltonian is that, generically, it has only topological degeneracies. The nontopological degeneracies that may arise in the SN model for noncommutative UFCs (see Ref.~\cite{Ritz24}) are indeed due to the very specific choice of couplings discussed above.

%
%
\section{Basic properties of degeneracies}
\label{sec:deg}
\begin{figure}[t]
\includegraphics[width=0.6 \columnwidth]{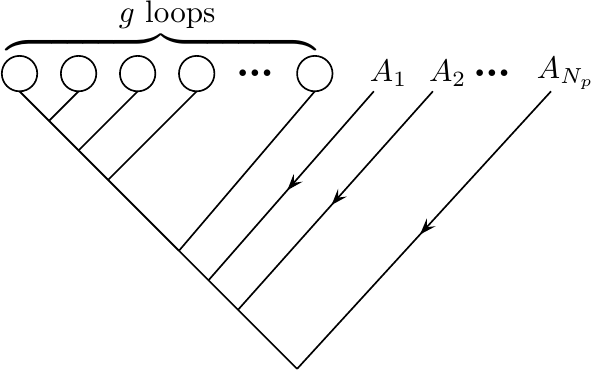}
\caption{Fusion tree indicating the topological degeneracy for $N_{\rm p}$ plaquettes having fluxons $A_1, \ldots, A_{N_{\rm p}}$ on a genus $g$ surface. For clarity, the internal multiplicity index $a_i$ associated with $A_i$'s and the fusion multiplicity index associated with each vertex are not displayed here.}
\label{fig:fusiontree0}
\end{figure}

%
%
In this section, we review some important results about the eigenstate topological degeneracies presented in Ref.~\cite{Ritz24} and extend these to the case of surfaces with punctures. Let us consider a RSN model defined on a trivalent graph with $N_{\rm  p}$ plaquettes embedded on an orientable closed surface of genus $g$. An eigenstate of $H$ defined in Eq.~\eqref{eq:RSN} can be represented by a fusion diagram (see Fig.~\ref{fig:fusiontree0}). In this picture,  each directed line is related to a fluxon $A_p$ with a multiplicity index $a_p$ and each vertex having three incident edges $A,B,C \in {\cal Z}({\cal C})$ is labeled with an integer (not shown) ranging from 1 to $N_{ABC}$ (labelings with $N_{ABC}= 0$ are not allowed), where $N_{ABC}$'s are the fusion multiplicity coefficients of the Drinfeld center. Such a fusion diagram implicitly assumes that the overall fusion product is the vacuum $\bf 1$. 

As discussed in Ref.~\cite{Ritz24}, for a given fluxon configuration \mbox{$\{A_1, A_2, \ldots, A_{N_{\rm p}}\} \in {\cal F} \subseteq {\cal Z}({\cal C})$}, with multiplicity indices \mbox{$\{a_1, a_2, \ldots, a_{N_{\rm p}}\}$}, the degeneracy of the  energy level, 
%
%
\be
E=- \sum_{p=1}^{N_{\rm p}} J_p^{A_p,a_p}, 
\label{eq:eigenenergy}
\ee
%
%
is given by counting all possible labelings of the unlabeled lines and vertices in Fig.~\ref{fig:fusiontree0}. It is important to note that this (topological) degeneracy only depends on the fusion rules of the Drinfeld center objects but not on the internal multiplicities. We emphasize that extra (nontopological) degeneracies may emerge for fine-tuned couplings as is the case for the SN Hamiltonian given in Eq.~\eqref{eq:ham}~\cite{Ritz24}. 

%
\subsection{Digression on fusion}
%

As explained above, an inwardly directed trivalent vertex $A,B,C$ in a fusion diagram such as  Fig.~\ref{fig:fusiontree0} is described by a quantum number ranging from 1 to $N_{ABC}$. One particular case is when one of the edges is already in the vacuum state $\bf 1$. Then, the two other edges must be in states $A$ and $\bar{A}$, where $\bar{A}$ is the dual of the object $A$, i.e., $N_{AB\bf 1}=\delta_{B,\bar{A}}$. Graphically, one can go from $A$ to $\bar{A}$ by reversing the orientation of the corresponding edge. Therefore, an equivalent notation for $N_{ABC}$ is $N_{AB}^{\bar{C}}$, where the emphasis is put on the fact that two incoming edges $A$ and $B$ fuse into an outwardly directed edge $\bar{C}$.

 In the following, we will use the notation
%
%
\beqn
&&N_{B_1 B_2 B_3 \ldots B_m}^X = \nonumber \\
&&\sum\limits_{\substack{Y_1, \ldots, Y_{m-2}\\ \in {\cal Z}({\cal C})}} N_{B_1 B_2}^{Y_1}  N_{Y_1 B_3}^{Y_2}  \ldots N_{Y_{m-3} B_{m-1}}^{Y_{m-2}}  N_{Y_{m-2} B_m}^X, \quad 
\label{eq:multh}
\eeqn
%
%
to notate the total number of ways that $m$ objects $B_1, \ldots, B_m$ can fuse to $X$.  This $N$ symbol is symmetric under exchange of any of the lower indices. We can also raise and lower indices as follows:
%
%
\be
N^X_{B_1 B_2 \ldots B_m} = N^{{\bf 1}}_{B_1 B_2 \ldots B_m \bar{X}} = N^{\bar B_1}_{\bar X B_2 \ldots B_m}.
\ee
%
%
Finally, we can paste together fusion trees with respective fusion outcomes $X$ and $\bar{X}$ by using the following relation:
%
%
\be
N^{{\bf 1}}_{B_1 B_2 \ldots B_m} = \sum_{X \in {\cal Z}({\cal C})} N^X_{B_1, \ldots, B_p} N^{\bar{X}}_{B_{p+1} \ldots B_m}.
\label{eq:generalizedN}
\ee
%
%

%
\subsection{Topological degeneracies}
%

Using the notation introduced in~\eqref{eq:multh}, we can compute the topological degeneracy of states for a genus $g$, $N_{\rm p}$ plaquettes system with fluxons $A_1, \ldots, A_{N_{\rm p}}$ through the plaquettes as: 
%
%
\beqn
&&\dim(g; A_1, \ldots, A_{N_{\rm p}}) =\nonumber \\ 
&&\sum_{B_1, \ldots B_g \in {\cal Z}({\cal C})} N_{B_1 \bar B_1 B_2 \bar B_2 \ldots B_g \bar B_g A_1 A_2 \ldots A_{N_{\rm p}}}^{\bf 1}.  \label{eq:multifusionN}
\eeqn
%
%
This form, which corresponds to the fusion tree of Fig.~\ref{fig:fusiontree0}, is given in Ref.~\cite{Ritz24} and is also re-derived in Appendix~\ref{app:surg}. 

We can greatly simplify expressions of this sort using the Verlinde formula~\cite{Verlinde88}
%
%
\be
N_{A, B}^C = \sum_{D \in   {\cal Z}({\cal C}) } \frac{S_{A, D}S_{B, D} S_{C, D}^*}{S_{\mathbf{1}, D}},
\label{Wq2}
\ee
%
%
where $S$ is the modular $S$-matrix of the Drinfeld center ${\cal Z}({\cal C})$.

Using this form along with the fact that $S$ is unitary and symmetric, one obtains the Moore-Seiberg-Banks result~\cite{Moore89}: 
%
%
\be
\dim(g; A_1, \ldots, A_{N_{\rm p}}) =  \sum_{C \in {\cal Z}({\cal C})} \!\!\left[ \prod_{p=1}^{N_{\rm p}}  S_{A_p,C} \right]  S_{\mathbf{1},C}^{2 - 2 g - N_{\rm p}} \label{eq:MooreSeiberg}.
\ee
%
%

%
\subsection{Degeneracies from punctured surfaces}
\label{sub:cut}
%

\begin{figure}[t]
    \includegraphics[width=3cm]{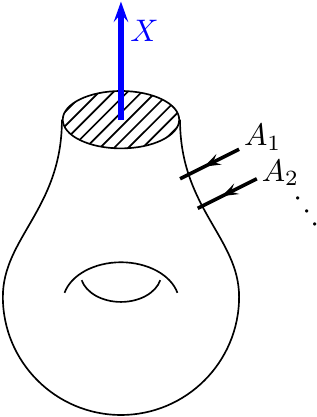}
    \caption{A genus $g$ surface with a disk (shaded) removed.  Here, we have drawn $g=1$ and we have allowed a quantum number $X \in {\cal Z}({\cal C})$ to be coming out of the hole.  In addition, we have shown that fluxons $A_p \in {\cal F}$ are penetrating the plaquettes on the surface. }
    \label{fig:genus1}
\end{figure}

We now  cut a disk out of the surface and calculate the effective dimension (degeneracy) of this partial surface, allowing for a particular quantum number of ``flux"  $X\in \mathcal{Z(C)}$, potentially different from the vacuum and \textit{not necessarily a fluxon} to be coming out of the hole   as shown in Fig.~\ref{fig:genus1}.   If this object has $N_{\rm p}$ plaquettes plus the disk-hole,  the topological degeneracy associated with it is just $\dim(g; A_1, \ldots, A_{N_{\rm p}}, \bar{X})$. In other words, the excitation through the removed disk is just like a  plaquette excitation, although we allow its quantum number to take any value $X \in \cal Z({\cal C})$ rather than forcing it to always be a fluxon.  Furthermore, since we intend to sew this object on to other similar objects such that the disk-hole is patched up, we do not include any internal degeneracy for the hole. Indeed, as explained in Ref.~\cite{Ritz24}, these internal degeneracies are only relevant for open strings.
Using Eqs.~\eqref{eq:multifusionN}-\eqref{eq:MooreSeiberg}, one then gets
%
%
\beqn
\dim(g; A_1, \ldots, A_{N_{\rm p}}, \bar{X}) & = &
 \sum_{C \in {\cal Z}({\cal C})}   \left[ \prod_{p=1}^{N_{\rm p}}  S_{A_p,C} \right]   \times \nonumber \\
&&   \:\: S_{\bar{X},C} \: \: S_{\mathbf{1},C}^{2 - 2 g - (N_{\rm p} +1)}. \label{eq:dimX}
\eeqn
%
%

%
%
%
%
Importantly, for $g=0$, $\dim(g; A_1, \ldots, A_{N_{\rm p}}, \bar{X})$ vanishes unless $X$ can be obtained by the fusion of multiple fluxons. We write $X \in {\cal F}^{\otimes}$, where $\mathcal{F}^\otimes$ denotes the subset of simple objects of $\mathcal{Z(C)}$ that can be obtained by fusion of fluxons (i.e., elements of ${\cal F}$). Hence, one has:   
%
%
\be
\mathcal{F} \subseteq \mathcal{F}^\otimes \subseteq \mathcal{Z(C)},
\label{eq:sets}
\ee
%
%
Similarly, for $g=1$, we need to have $X \in ({\cal F}^{\otimes} \times (B_1 \times \bar B_1))$ for some object $B_1\in \mathcal{Z(C)}$ around the handle, and so forth.  More generally, $\dim(g; A_1, \ldots, A_{N_{\rm p}},\bar{X})$ vanishes if  $X \notin {\cal F}^{\otimes} \times {\cal G}^{\otimes g}$, where $\cal G$ is the set of objects in the fusion products $B \times \bar{B}$, for all $B \in {\cal Z}({\cal C})$, and ${\cal G}^{\otimes g}$ means the set of objects generated  by fusing $g$ objects in $\cal G$ together.  
%
%
\begin{figure}
\includegraphics[width=1in,angle=180]{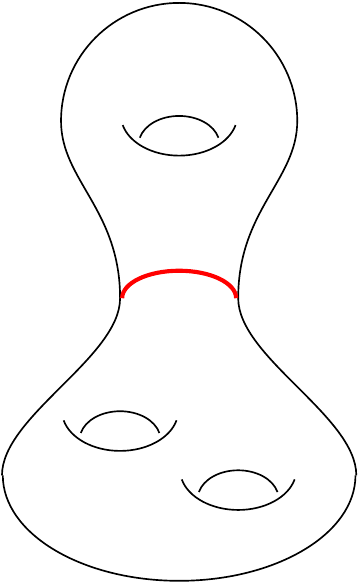}
\caption{Sewing together two objects as in Fig.~\ref{fig:genus1}.  Here, we have sewn together genus $g_{\mathcal{R}}=1$ (lower half) with $g_{\overline{\mathcal{R}}}=2$ (upper half), connected at the red line, to obtain an object with genus $g=g_{\mathcal{R}}+ g_{\overline{\mathcal{R}}} = 3$.  We say that the red line here goes around a {\it throat} separating $g_{\mathcal{R}}=1$ from $g_{\overline{\mathcal{R}}}=2$.}
\label{fig:sewg1g2}
\end{figure}
%
%

These effective dimensions are arranged so that we can sew together two of these surfaces at their holes to obtain the degeneracy of a surface with no holes, as shown in Fig.~\ref{fig:sewg1g2}.  The procedure of cutting disk holes in surfaces and reconnecting them in this way is known as taking the ``connected sum" of surfaces.  In particular,  we want to assemble together such an object $\mathcal{R}$ of genus $g_\mathcal{R}$ with $N_{\rm p}^\mathcal{R}$ plaquettes having fluxes $A_1, \ldots, A_{N_{\rm p}^\mathcal{R}}$ and a flux $X$ coming out of the hole, together with an object $\overline{\mathcal{R}}$ of genus $g_{\overline{\mathcal{R}}}$ with $N_{\rm p}^{\overline{\mathcal{R}}}$ plaquettes having fluxes $A_{{N_{\rm p}^\mathcal{R}}+1} \ldots A_{N_{\rm p}}$ with flux $X$ going   into the hole ($\bar X$ coming out of the hole).  The resulting object $\mathcal{R} \cup \overline{\mathcal{R}}$ genus $g=g_\mathcal{R} + g_{\overline{\mathcal{R}}}$ surface, as shown in Fig.~\ref{fig:sewg1g2}, then has a total effective  dimension [making use of the property of Eq.~\eqref{eq:generalizedN}]
%
%
\beqn
\dim(g; A_1, \ldots, A_{N_{\rm p}}) = && \!\!\!\!\!\!\sum_{X \in {\cal Z}({\cal C})} \!\!\!  \dim(g_\mathcal{R}; A_1, \dots, A_{N_{\rm p}^\mathcal{R}},\bar{X}) \nonumber \\ 
&&\dim(g_{\overline{\mathcal{R}}}; A_{{N_{\rm p}^\mathcal{R}}+1}, \dots, A_{N_{\rm p}},X). \nonumber \\ 
\label{eq:dimXdimX}
\eeqn
%
%
%

%
%
\begin{figure}
\includegraphics[width=4cm]{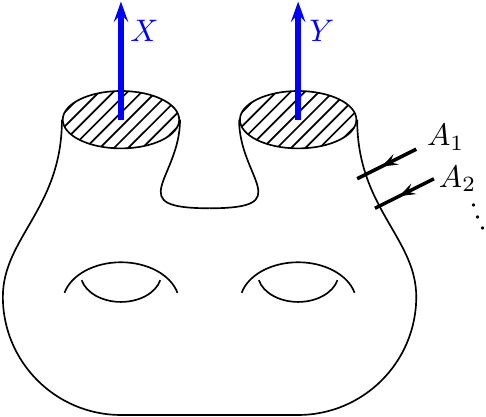}
\caption{A genus-$g$ surface with two disks (shaded) removed.  Here, we have drawn $g=2$ and we have allowed a quantum numbers $X, Y \in {\cal Z}({\cal C})$ to be coming out of the holes.  In addition we have shown that fluxons $A_j \in {\cal F}$ are penetrating the plaquettes on the surface.}
\label{fig:twocuts}
\end{figure}
\begin{figure}[t]
\includegraphics[width=1.2in,angle=180]{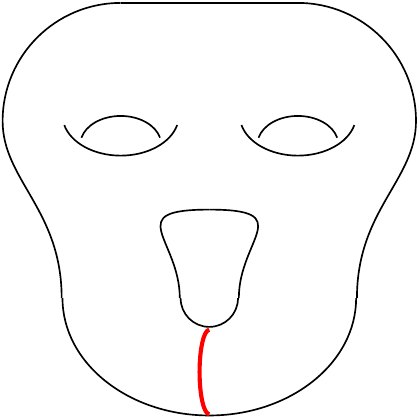}
\caption{Sewing together the two open faces of Fig.~\ref{fig:twocuts}.  The object in Fig.~\ref{fig:twocuts} is genus $g=2$, but when we sew together the two holes (which is only possible if $Y=\bar{X}$), we have an object of genus $g=3$.  The red line here goes around a {\it handle}.}
\label{fig:fidget}
\end{figure}

%
%
Similarly, we can write a degeneracy for a genus-$g$ surface with two disks removed having quantum numbers $X$ and $Y$ coming out of the missing disk-holes, as shown in Fig.~\ref{fig:twocuts}.   Using Eqs.~\eqref{eq:multifusionN}-\eqref{eq:MooreSeiberg},  the topological degeneracy of this object can be written as:
%
%
\beqn
\label{eq:dimXY}
\dim(g; A_1, \ldots, A_{N_{\rm p}}, \bar{X},\bar{Y})  = \!\!\!\!\!\!
&& \sum_{C \in {\cal Z}({\cal C})}   \left[ \prod_{p=1}^{N_{\rm p}}  S_{A_p,C} \right]   \times \nonumber \\
&&   \:\: S_{\bar{X},C} \: \: S_{\bar{Y},C} \: \: S_{\mathbf{1},C}^{2 - 2 g - (N_{\rm p} +2)}. \nonumber \\ 
\eeqn
%
%
Again, we have not included the internal degeneracy associated with the quantum number in the hole, since we intend to sew these holes together.  Setting $\bar{Y}=X$ in Eq.~\eqref{eq:dimXY}, we obtain the effective degeneracy for a surface with an additional handle, and a particular flux $X$ through that handle (see Fig.~\ref{fig:fidget}). For example, a sphere with two holes cut is a cylinder.  If we connect these two holes to each other, we get a torus.  More generally, if we cut two holes in a genus-$g$ object with plaquette fluxes $A_1, \ldots, A_{N_{\rm p}}$ and reconnect the two holes together we get a genus $(g+1)$ object. 
 The full topological degeneracy of such an object (with unrestricted flux through the handle) is then
%
%
\beqn
&&  \dim(g+1; A_1, \ldots, A_{N_{\rm p}}) = \sum_{X \in {\cal Z}({\cal C})} \label{eq:increasegenus}\\ 
&&   \dim(g; A_1, \ldots, A_{N_{\rm p}}, \bar{X},X).\nonumber
\eeqn
%
%
An example of this type of sewing is shown in Fig.~\ref{fig:fidget}.
%
%
%
%

%
%
\section{Partition functions}
\label{sec:partfunc}
%
%
%
\subsection{Results for the RSN model}
\label{sub:ZRSN}
%
Once the degeneracies of the energy levels are determined, it is pretty easy to compute the finite-temperature partition function of the RSN model \eqref{eq:RSN}. For a genus-$g$ surface with $N_{\rm p}$ plaquettes, it is given by 
%
%
\beqn
Z(g, N_{\rm p}) &=& {\rm Tr}({\rm e}^{-\beta H}) \nonumber\\ 
&=& \sum_{A_1,  \ldots, A_{N_{\rm p}} \in {\cal F}}      \dim(g; A_1, \ldots, A_{N_{\rm p}}) \nonumber \\
&&   \sum_{a_1=1}^{n_{{A_1},1}} \:\sum_{a_2=1}^{n_{{A_2},1}} \ldots \sum_{a_{N_{\rm p}}=1}^{n_{{A_{N_{\rm p}}},1}} {\rm e}^{\beta \sum_{p=1}^{N_{\rm p}} J_p^{A_p,a_p}}, \label{eq:part0}
\eeqn
%
%
where $\beta=1/T$, is the inverse temperature (we set \mbox{$k_B=1$}), and where the energy associated with a fixed fluxon configuration of \mbox{$\{A_1, A_2, \ldots, A_{N_{\rm p}}\}$}, with multiplicity indices \mbox{$\{a_1, a_2, \ldots, a_{N_{\rm p}}\}$}, is given by Eq.~\eqref{eq:eigenenergy}.
Using Eq.~\eqref{eq:MooreSeiberg}, and the fact that $S$ is a symmetric and unitary matrix, one can rewrite this partition function in the following simple form:
%
%
\be
\label{eq:ZRSN}
Z(g,N_{\rm p}) =  
\sum_{C \in {\cal Z}({\cal C})} S_{{\bf 1},C}^{2 - 2g} \prod_{p=1}^{N_{\rm p}} z_{p,C}, 
\ee
%
%
where we introduced
%
%
\be
\label{eq:zpc}
z_{p,C} =\sum_{A \in {\cal F}} \sum_{a=1}^{n_{A,1}}\frac{S_{A, C}}{S_{{\bf 1},C}} \, {\rm e}^{\beta J_p^{A,a}}.
\ee
%
%
It is interesting to analyze the global structure of Eq.~\eqref{eq:ZRSN}, which is one of the main results of our paper. Indeed, this partition function is written as a sum of terms which are simple products over all plaquettes.  This is reminiscent of the fact that the Hamiltonian is a sum of local commuting projectors together with some nonlocal constraints given by the topological degeneracies. 

Interestingly, in the thermodynamic limit, the sum in Eq.~\eqref{eq:ZRSN} is dominated by a special set of objects $C$ which maximizes the ratio $S_{A, C}/S_{{\bf 1},C}$. As explained in Appendix~\ref{app:transpflu}, for  $A\in \mathcal{F}$, this quantity is maximized if $C$ belongs to the pure fluxon set $\mathcal{P}\subseteq \mathcal{F}$. A pure fluxon $C$ is a fluxon which obeys
$n_{C,1}=d_C$. On the one hand, according to Eq.~\eqref{eq:SABform}, for any $A$ and $C \in \mathcal{Z(C)}$, we have 
%
%
\be
|S_{A,C}| \leq d_A d_C/{\cal D},
\label{ineq}
\ee
%
%
where $d_A$ is the quantum dimension of the particle $A$ and 
%
%
\be
\mathcal{D}=\sqrt{\sum_{C \in \mathcal{Z}(\mathcal{C})} d_C^2},
\ee
%
%
is the total quantum dimension of $\mathcal{Z}(\mathcal{C})$. On the other hand, if $A \in \mathcal{F}$ and  $C\in \mathcal{P}$ (or the contrary), the inequality \eqref{ineq} becomes an equality \mbox{$S_{A,C}= d_A d_C/{\cal D}$}. In other words, {\it pure fluxons braid trivially with all fluxons}.
Hence, when $C$ is a pure fluxon, one has
%
%
\be
z_{p,C} = \sum_{A \in {\cal F}} \sum_{a=1}^{n_{A,1}} d_A   {\rm e}^{\beta J_p^{A,a}}\equiv z_p,
\label{eq:Ipureflux}
\ee
%
%
is independent of $C$. For any $T>0$. we therefore get
%
%
\be
Z (g,N_{\rm p}) \underset{N_{\rm p} \rightarrow \infty}{\simeq} M_g \,\, \mathcal{D}^{2g-2}  \prod_{p=1}^{N_{\rm p}} z_{p},
\label{eq:ZthermoRSN}
\ee 
%
%
where we introduced
%
%
\be
M_g \equiv  \sum_{A \in \mathcal{P}} n_{A,1}^{2 - 2g}= \sum_{A \in \mathcal{P}} d_A^{2-2g} .
\label{M}
\ee
%
%
For a commutative input category, $M_g$ is simply the number of pure fluxons as $n_{A,1}=1$ for $A\in \mathcal{F}$, i.e., $M_g=|\mathcal{P}|$, for any genus. It also corresponds to the number of Abelian fluxons.

%
\subsection{Infinite-temperature limit and Hilbert-space dimension}
\label{sec:hilbpure}
%

The Hilbert-space dimension can be easily computed by taking the infinite-temperature limit of the partition function\eqref{eq:ZRSN}. In this limit, using the fluxon identity 
%
%
\be
S {\bf n}_1={\bf n}_1,
\label{eq:fluxonid}
\ee
%
%
where ${\bf n}_1$ is the vector with components $n_{C,1}$ ($n_{C,1}=0$, if $C\notin  {\cal F}$)~\cite{Ritz24}, one gets
%
%
\be
\lim_{T \rightarrow \infty} z_{p,C} =\frac{n_{C,1}}{S_{{\bf 1}, C}}.
\label{eq:zpcthermo}
\ee
%
%
Then, one straightforwardly obtains:
%
%
\beqn
\dim \mathcal{H} &=& \lim_{T \rightarrow \infty}Z(g, N_{\rm p}), \nonumber \\
&=&\sum_{C \in {\cal Z}({\cal C})} S_{{\bf 1}, C}^{2 - 2g} \left(\frac{n_{C,1}}{S_{{\bf 1}, C}} \right) ^{N_{\rm p}}, \nonumber\\
&=&\sum_{C \in {\cal F}} S_{{\bf 1}, C}^{-N_\text{v}/2} n_{C,1}^{N_{\rm p}}.
\eeqn
%
%
For commutative input ${\cal C}$, this result is in agreement with the one given in Ref.~\cite{Vidal22}. We also used the fact that, for a trivalent graph, the Euler-Poincar\'e characteristic gives 
%
%
\be
2-2g-N_{\rm p}=- N_\text{v}/2,
\label{eq:EPchar}
\ee
%
%
where $N_\text{v}$ is the number of vertices.

In the limit of large $N_{\rm p}$, we have 
%
%
\be
\dim \mathcal{H}  \underset{N_{\rm p} \rightarrow \infty}{\simeq} M_g \,\,\,  \mathcal{D}^{N_{\rm v}/2},
\label{dimHM}
\ee
%
%
which show that the Hilbert space is not a tensor product of local Hilbert spaces.

%
\subsection{Cutting and gluing partition functions}
\label{sub:cutgluepart}
%
Using our results on effective degeneracies of Sec.~\ref{sub:cut}, we can also write effective partition functions for surfaces with holes cut in them. Note that we are not including any energy associated with the flux through the hole.  For a genus-$g$ surface with a single hole carrying a flux $X\in \mathcal{Z(C)}$ (see Fig.~\ref{fig:genus1} for $g=1$) and $N_{\rm p}$ plaquettes, the (effective) partition function can be computed using Eq.~\eqref{eq:dimX} and reads as
%
%
\be
\label{eq:ZXsimp}
Z_X(g,N_{\rm p}) =  \sum_{C \in {\cal Z}({\cal C})} S_{\bar{X},C} S_{{\bf 1},C}^{1 - 2g}\prod_{p=1}^{N_{\rm p}} z_{p,C}.
\ee
%
%
where $z_{p,C}$ is given in Eq.~\eqref{eq:zpc}.

Analogous to what is done in Eq.~\eqref{eq:dimXdimX}, we can compute the partition function of  a surface obtained by gluing together one object of genus $g_\mathcal{R}$ with  $N_{\rm p}^{\mathcal{R}}$ plaquettes with another  object of genus $g_{\overline{\mathcal{R}}}$ with  $N_{\rm p}^{\overline{\mathcal{R}}}$ plaquettes
%
%
\be
Z(g_\mathcal{R} +  g_{\overline{\mathcal{R}}},N_{\rm p}^{\mathcal{R}}  + N_{\rm p}^{\overline{\mathcal{R}}})=\sum_{X \in {\cal Z}({\cal C})} \!\!\!\!\! Z_X(g_\mathcal{R},N_{\rm p}^{\mathcal{R}})  \, Z_{\bar X}(g_{\overline{\mathcal{R}}}, N_{\rm p}^{\overline{\mathcal{R}}}). 
\label{eq:ZXXbar}
\ee
%
%
This is precisely the sewing operation depicted in Fig.~\ref{fig:sewg1g2}.

Similarly, we can write an effective partition function associated to a surface with two holes in it (see Fig.~\ref{fig:twocuts}).
Again, using Eq.~\eqref{eq:dimXY}, this expression can be written as
%
%
\be
Z_{X,Y}(g,N_{\rm p})= \sum_{C \in {\cal Z}({\cal C})} S_{\bar{X},C} S_{\bar{Y},C} S_{{\bf 1},C}^{\:\:- 2g} \prod_{p=1}^{N_{\rm p}} z_{p,C}.
\label{eq:ZXYsimp}
\ee
%
%
These holes can then be glued to each other in order to increase the genus, as in Eq.~\eqref{eq:increasegenus}, to give
%
%
\be
Z(g+1,N_{\rm p})=\sum_{X \in {\cal Z}({\cal C})} Z_{X, \bar X}(g,N_{\rm p}),
\label{eq:handledec}
\ee
%
%
which is the sewing depicted in Fig.~\ref{fig:fidget}.

The three partition functions $Z$, $Z_X$, and $Z_{X,Y}$ will be used in Sec.~\ref{sec:proj} to compute the thermal averages of some operators. 

%
\section{Thermodynamics for the SN model}
\label{sec:SNmodel}
%
The simple form of the general partition function given in Eq.~\eqref{eq:ZRSN} allows one to study several quantities. Here, we shall discuss them for the SN model for which \mbox{$J_p^{A,a}=\delta_{A,\bf{1}}$}, for all $p$'s. In this case and keeping in mind that  $d_{\bf 1}=n_{{\bf 1},1}=1$, Eqs.~\eqref{eq:zpc}-\eqref{eq:fluxonid} lead to 
%
%
\be
z_{p,C}=\frac{n_{C,1}}{S_{{\bf 1},C}}-1+{\rm e}^{\beta},
\label{eq:zpCSN}
\ee
%
%
which is independent of $p$.  This result is particularly pleasing since it depends only on the internal multiplicities $n_{C,1}$ and on $S_{{\bf 1}, C}=d_C/\mathcal{D}$.
%
\subsection{Partition functions in the thermodynamic limit}
\label{sec:Z_th}
%
As explained in Sec.~\ref{sub:ZRSN}, in the large-$N_{\rm p}$ limit, the sum in Eq.\eqref{eq:ZRSN} is dominated by pure fluxons ($n_{C,1}=d_C$) so that, for the SN model, one has: 

%
%
\be
Z (g,N_{\rm p}) \underset{N_{\rm p} \rightarrow \infty}{\simeq} \left({\mathcal D} -1 + {\rm e}^{\beta}\right)^{{N_{\rm p}}}\sum_{C\in \cal{P}} S_{{\bf 1}, C}^{2-2g},
\label{eq:ZSNthermo}
\ee
%
%
where we used the identity $S_{{\bf 1},C}=d_C/\mathcal{D}$, for all $C$.  A similar result was obtained for the case of the Fibonaccci input category by~\cite{Hu14}. Here, we obtain it for any input UFC. 
Apart from a global factor  $M_g \mathcal{D}^{2g-2}$, the SN partition function \textit{in the thermodynamic limit} is the same as that for $N_\text{p}$ independent spins (with $q=\mathcal{D}$ states) in a magnetic field. As discussed in  Appendix~\ref{app:potts}, corrections to the thermodynamic limit show a relation to the 1D Potts model.
 
Similarly, one has: 
 
%
%
\be
Z_X(g,N_{\rm p})  \underset{N_{\rm p} \rightarrow \infty}{\simeq} \left({\mathcal D} -1 + {\rm e}^{\beta}\right)^{{N_{\rm p}}}\sum_{C\in \cal{P}} S_{\bar{X},C} S_{{\bf 1}, C}^{1-2g},
\label{eq:ZXSNthermo}
\ee
%
%
and 
%
%
\be
Z_{X,Y} (g,N_{\rm p})  \underset{N_{\rm p} \rightarrow \infty}{\simeq} \left({\mathcal D} -1 + {\rm e}^{\beta}\right)^{{N_{\rm p}}} \sum_{C\in \cal{P}} S_{\bar{X},C} S_{\bar{Y},C}S_{{\bf 1}, C}^{-2g}.
\ee
%
%

%
\subsection{Ground-state degeneracy}
\label{sec:zeroT}
%
The zero-$T$ limit of  the partition function $Z$ allows one to compute straightforwardly the ground-state degeneracy of the SN Hamiltonian  \eqref{eq:ham}. This topology-dependent degeneracy is given by: 
%
%
\be 
 \lim_{T \rightarrow 0} Z \:{\rm e}^{\beta E_0}=  \sum_{C \in {\cal Z}({\cal C})}  S_{{\bf 1},C}^{2 - 2g},
   \label{eq:degenT} 
 \ee
%
%
where $E_0=-N_{\rm p}$, is the ground-state energy. This result is exactly the topological quantum field theory (TQFT) result~\cite{Moore89} for the degeneracy on a surface of genus $g$ that can also be obtained directly from Eq.~\eqref{eq:MooreSeiberg} by imposing the no-flux condition,  $A_p={\bf 1}$, for all $p$'s, which defines the ground-state manifold of the SN model.

%
%
\subsection{Energy, specific heat, and entropy}
\label{sec:thermo}
%
%
Using Eqs.~\eqref{eq:ZRSN} and \eqref{eq:zpCSN}, it is straightforward to compute several thermodynamical quantities for the SN model. Here, we discuss some of them by considering directly the thermodynamic limit \eqref{eq:ZSNthermo} for which expression becomes especially simple. 

In the large-$N_{\rm p}$ limit, the energy per plaquette is given by
%
%
\be
e =  \lim_{N_{\rm p} \rightarrow \infty}- \frac{1}{N_{\mathrm p}} \frac{\partial \ln Z }{\partial \beta}=-\, \frac{\mathrm{e}^{\beta} }{\mathcal{D}-1+\mathrm{e}^{\beta}}.
\label{eq:nrj}
\ee
%
%
We will see in Sec.~\ref{sub:contra} how to obtain this result from the projectors directly [see Eq.~\eqref{eq:nrjproj}].
  
Similarly, the specific heat per plaquette is given by
%
%
\be
c =\lim_{N_{\rm p} \rightarrow \infty} \frac{\beta^2}{N_{\mathrm p}}\frac{\partial^2 \ln Z}{\partial \beta^2}= \frac{ \mathrm{e}^{\beta} \, \beta^2 \,(\mathcal{D}-1)}{(\mathcal{D}-1+\mathrm{e}^{\beta})^2}.
\ee
%
%

These expressions which hold for any UFC are exactly the same as the one derived in Ref.~\cite{Vidal22} where only modular input UFCs were considered.  This very simple form of the specific heat shows that, $c$ is always a smooth function of the temperature, indicating the {\it absence of finite-temperature phase transition in this model}. This latter result also holds for the RSN model and is in agreement with the general result derived by Hastings~\cite{Hastings11} which holds for any local commuting projector Hamiltonian in two dimensions. The specific heat features a maximum known as a Schottky anomaly which is typical of independent spins as well as interacting spin chains in the thermodynamic limit (see, e.g., Ref~\cite{Sator_book}).

Finally, the entropy is:
%
%
\be
S=- \beta \frac{\partial \ln Z}{\partial \beta} + \ln Z.
\label{eq:entropy}
\ee
%
%
Using Eq.~\eqref{eq:ZSNthermo}, one gets in the thermodynamic limit at $T>0$ 
%
%
\be
S \underset{N_{\rm p} \rightarrow \infty}{\simeq} N_{\rm p} \left[\ln(\mathcal{D}-1+e^{\beta})-\frac{\beta e^{\beta}}{\mathcal{D}-1+e^{\beta}} \right] +\ln \frac{M_g}{\mathcal{D}^{2 -2g}},
\label{eq:entropy2}
\ee
%
%
where  $M_g$ is given in Eq.~\eqref{M}. This consists of a volume (extensive) term and a constant term. In the infinite-temperature limit, this expression simply becomes
%
%
\be
\lim_{T \to \infty} S =\ln (\dim \mathcal{H}) \underset{N_{\rm p} \rightarrow \infty}{\simeq}  \frac{N_{\rm v}}{2} \ln \mathcal{D}+\ln M_g.
\label{eq:enttot}
\ee
%
%
For future reference, we note that the constant term, $\ln M_g$, is not related to quantum entanglement but to the fact that the Hilbert space is constrained by the fusion rules (vertex defects are forbidden).

%
%
\section{Thermal average of projectors}
\label{sec:proj}
%
%
Having established the basic thermodynamics, we turn to calculate the thermal expectation of certain operators of interest. The first operators we will study are the topological projection  operators $P_{X}(L)$ for $X \in {\cal Z}({\cal C})$ and $L$ a closed path on our surface. Following the same line as in Sec.~\ref{sub:cut}, we will keep in mind that $L$ is a closed contour separating two regions $\mathcal{R}$ and $\overline{\mathcal{R}}$.

\begin{figure}[t]
\includegraphics[width=0.4 \columnwidth]{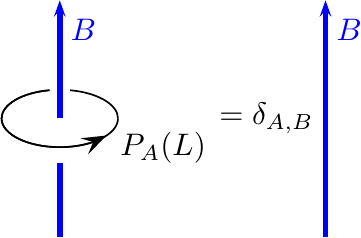}
\caption{Schematic representation of a  projector $P_A$ measuring a particle $B$.  The loop $L$ goes counterclockwise around the region being projected.}
\label{proj}
\end{figure}

As a complete set of orthogonal projection operators, these satisfy
%
%
\be
P_X(L) P_Y(L) = \delta_{X,Y} P_X(L),
\label{eq:PP0}
\ee
%
%
and 
%
%
\be
\label{eq:Pidid}
\sum_{X \in {\cal Z({\cal C})}}  P_X(L) = \mathds{1},
\ee
%
%
where the sum is over all objects of the Drinfeld center $Z({\cal C})$.  Physically, $P_{X}(L)$ projects to a configuration where the flux through the loop $L$ is given by the particle type $X \in \mathcal{Z}(\mathcal{C})$.  For definiteness, we define the direction of the loop $L$ such that it travels counterclockwise around the region of interest.  This projection is shown graphically in Fig.~\ref{proj}.  The expectation of these operators will be  obtained using our cutting and gluing rules from Sec.~\ref{sub:cutgluepart}.
We consider here three different types of loops.  Firstly, we discuss loops around a {\it handle} $L_h$,  (see red line in Fig.~\ref{fig:fidget}), and secondly  loops around a {\it throat} $L_t$ (see red line in Fig.~\ref{fig:sewg1g2}). The case of a {\it contractible} loop $L_c$ is finally discussed as a special case of throat. The main results of this section are given in Eqs.~\eqref{eq:exphand}, \eqref{eq:expth}, and \eqref{eq:Pcont}. 

Remarkably, these expressions are exact for any temperature, any system size, any trivalent graph, any choice of the couplings, and any input UFC.

%
%
%
%

%
%
\subsection{Topological expectations}
%
%

Depending on the couplings $J_p^{A,a}$'s, the ground state of the Hamiltonian \eqref{eq:RSN} may or may not have nontrivial
fluxons through plaquettes.  At zero temperature, our projectors will measure the topological quantum numbers [i.e. $X\in \mathcal{Z(C)}$] through these loops --- which are fixed by the nature of the ground state.

At any finite temperature, there is a nonzero probability that each of the plaquettes will not be in their ground state.  These excitations carry topological quantum numbers, and we should expect that, for a sufficiently large system, the topological properties of the system will be scrambled.  For example, a loop encircling a large enough region will surround a completely unknown topological quantum number. However, not all topological information will be erased at finite temperature at long distances.  Since we have enforced the vertex constraint at the level of the Hilbert space (rather than as a term in the Hamiltonian as was originally done in Ref.~\cite{Levin05}), we have forbidden certain types of defects, and topological information associated to them can  remain at any temperature.

%
%
\subsection{Handles}
\label{sub:handles}
%
%
To implement a projector on a loop $L_h$ around a handle of a genus-$g$ surface with $N_{\rm p}$ plaquettes, we refer back to Eq.~\eqref{eq:handledec}.  This representation of the partition function of the full system is already written in terms of a sum of effective partition function $Z_{X,\bar{X}}$ where a flux $X$ is imposed in  the chosen handle after gluing.  Thus, the thermal average of the operator that projects to have flux $X$ through the chosen handle is simply given by 
%
%
\be
\langle P_X(L_h) \rangle = \frac{Z_{X,\bar X}(g -1, N_{\rm p})}{Z(g, N_{\rm p})}, 
\label{eq:exphand}
\ee
%
%
where $Z$ and $Z_{X,\bar{X}}$ are given for arbitrary couplings in Eqs.~\eqref{eq:ZRSN} and \eqref{eq:ZXYsimp}, respectively. This quantity, which only depends on $\beta$, $g$ and $N_{\rm p}$, vanishes if $X \notin {\cal F}^{\otimes} \times {\cal G}^{\otimes g}$ [see discussion after Eq.~\eqref{eq:sets}]. Using Eq.~\eqref{eq:handledec}, one can easily check the identity \eqref{eq:Pidid}.

As done for the partition functions, it is interesting to study the behavior of $\langle P_X(L_h) \rangle$ in some limiting cases. Let us first discuss the large-$N_{\rm p}$ limit at {\it any} nonzero temperature.  In this limit, as explained in Sec.~\ref{sub:ZRSN}, the sums over all objects of the Drinfeld centers  appearing in Eq.~\eqref{eq:exphand} are dominated by the pure fluxons, and $z_{p,C}$ becomes independent of $C$ [see Eq.~\eqref{eq:Ipureflux}]. Hence, one simply obtains:
%
%
\be
\lim_{N_{\rm p} \rightarrow \infty} \langle P_X(L_h) \rangle = \frac{\sum\limits_{C \in {\cal P}} S_{X,C}  S_{\bar{X},C}\, S_{{\bf 1},C} ^{2-2g}} {\sum\limits_{C \in {\cal P}} S_{{\bf 1},C} ^{2-2g}}.
\label{eq:Mratg1}
\ee
%
%
Again, unitarity of the symmetric $S$-matrix ensures the relation \eqref{eq:Pidid}. The result of Eq.~\eqref{eq:Mratg1}  is independent of the couplings and of the temperature, as long as $T>0$.  

One can also easily get the behavior of $\langle P_X(L_h) \rangle$ in the infinite-temperature limit. In this limit, using Eq.~\eqref{eq:zpcthermo}, Eq.~\eqref{eq:exphand} becomes
%
%
\be
\lim_{T \rightarrow \infty} \langle P_X(L_h) \rangle = \frac{\sum\limits_{C \in {\cal Z}({\cal C})} S_{X, C} S_{\bar X, C}  \, S_{{\bf 1}, C}^{2-2g - N_{\rm p}} n_{C,1}^{N_{\rm p}}} {\sum\limits_{C \in {\cal Z}({\cal C})}  \, S_{{\bf 1}, C}^{2-2g -N_{\rm p}} n_{C,1}^{N_{\rm p}} },
\ee
%
%
which is nontrivial because we are considering a restricted Hilbert space (no violation of the vertex constraints). Remarkably, this result is also independent of the couplings $J_p^{A,a}$.\\

Finally, as for the zero-$T$ partition function, $\langle P_X(L_h) \rangle$ depends on the couplings. For the SN model, one gets
%
%
\be
\lim_{T \rightarrow 0}\langle P_X(L_h) \rangle = \frac{\sum\limits_{C \in {\cal Z}({\cal C})} S_{X,C}  S_{\bar{X},C} S_{{\bf 1}, C}^{2 - 2 g}}{\sum\limits_{C \in {\cal Z}({\cal C})} S_{{\bf 1}, C}^{2 - 2g } },
\label{eq:PhandleSN}
\ee
%
%
which is valid for any value of $N_{\rm p}$.  This result differs from  Eq.~\eqref{eq:Mratg1} in that the sum is over all objects of the Drinfeld center ${\cal Z}({\cal C})$ rather than just the subset of pure fluxons ${\cal P}$. 
 
%
%
\subsection{Throats}
\label{sub:throats}
%
%
Let us now discuss the case where the loop is defined around a throat, i.e., a loop $L_t$ around a contour separating genus $g_\mathcal{R}$ and $g_{\overline{\mathcal{R}}}$ surfaces such as the red line shown in Fig.~\ref{fig:sewg1g2}.  To compute $\langle P_X(L_t) \rangle$, we follow roughly the same strategy, writing the genus-$g=g_\mathcal{R}+g_{\overline{\mathcal{R}}}$ partition function in terms of partition functions of the two pieces as in Eq.~\eqref{eq:ZXXbar}.  Fixing the value of the flux though the throat to $X$ gives us
%
%
\be
 \langle P_X(L_t) \rangle = \frac{Z_{X}(g_\mathcal{R}, N_{\rm p}^{\mathcal{R}})Z_{\bar X}(g_{\overline{\mathcal{R}}}, N_{\rm p}^{\overline{\mathcal{R}}})}{Z(g, N_{\rm p})},
 \label{eq:expth}
\ee
%
%
where $N_{\rm p}=N_{\rm p}^{\mathcal{R}} + N_{\rm p}^{\overline{\mathcal{R}}}$ is the total number of plaquettes. 
Again, one can easily check the identity \eqref{eq:Pidid} using Eq.~\eqref{eq:ZXXbar}. 

As done for $\langle P_X(L_{h}) \rangle$, it is straightforward to extract various limiting cases (such as large $N_{\rm p}^{\mathcal{R}}$, large $N_{\rm p}^{\overline{\mathcal{R}}}$, or infinite-temperature limits) using Eqs.~\eqref{eq:Ipureflux} and \eqref{eq:zpcthermo}.

%
%
\subsection{Contractible loops}
\label{sub:contra}
%
%
To conclude this section, we consider the case of contractible loops $L_c$. As one can observe in Fig.~\ref{fig:sewg1g2}, if $g_\mathcal{R}=0$ or $g_{\overline{\mathcal{R}}}=0$, the red line becomes contractible. Thus, $\langle P_X(L_{c})\rangle$  can be directly computed from Eq.~\eqref{eq:expth} by setting one of the genuses, let us say $g_{\overline{\mathcal{R}}}$,  to 0. We then have
%
%
\be
\langle P_X(L_c) \rangle= \frac{Z_X (g_\mathcal{R}, N_{\rm p}^{\mathcal{R}})Z_{\bar{X}} (0, N_{\rm p}^{\overline{\mathcal{R}}})}{Z(g, N_{\rm p})},
\label{eq:Pcont}
\ee
%
%
where the contractible side has $N_{\rm p}^{\overline{\mathcal{R}}}$ plaquettes, while the other side has genus $g_\mathcal{R}$ and $N_{\rm p}^{\mathcal{R}}$ plaquettes, and the full system has genus $g=g_\mathcal{R}$ and $N_{\rm p}=N_{\rm p}^{\mathcal{R}}+N_{\rm p}^{\overline{\mathcal{R}}}$ plaquettes.  Because $g_{\overline{\mathcal{R}}}=0$, this expectation value must be zero unless $X \in {\cal F}^{\otimes}$ [see also the comment just after Eq.~\eqref{eq:dimX}].  That is, $X$ must be obtainable by the fusion of multiple fluxons.\\

For the special case of the SN model, Eq.~\eqref{eq:Pcont} becomes very simple. For instance, in the zero-$T$ limit, one can check that
%
%
\be
\lim_{T \rightarrow 0}\langle P_X(L_c) \rangle =\delta_{X,{\bf 1}},
\label{eq:zeroTPSN}
\ee
%
%
which simply indicates that, for all ground states of the SN model, there is no nontrivial fluxon in any contractible loop. Noncontractible loops may, however, contain nontrivial fluxes [see, e.g., Eq.~\eqref{eq:PhandleSN} for handles].

In the limit where the side with genus $g_\mathcal{R}$ is large ($N_{\rm p}^{\mathcal{R}}\gg 1$) using  Eqs.~\eqref{eq:zpCSN}, \eqref{eq:ZSNthermo}, and \eqref{eq:ZXSNthermo}, one gets: 
%
%
\beqn
\lim_{N_{\rm p}^{\mathcal{R}} \rightarrow \infty}  \langle P_X(L_c) \rangle &=& d_X \sum_{C \in {\cal Z}({\cal C})} S_{\bar X,C} S_{{\bf 1},C} \times \nonumber \\
&& \left(\frac{{\cal D} \frac{n_{C,1}}{d_C}-1+{\rm e}^{\beta}}{{\cal D} - 1 + {\rm e}^\beta}   \right)^{N_{\rm p}^{\overline{\mathcal{R}}}},
\label{eq:50}
\eeqn
%
%
if $X \in {\cal F}^{\otimes}$, and 0 otherwise. 

In the special case where $X={\bf 1}$ and $N_{\rm p}^{\overline{\mathcal{R}}} = 1$, the projection operator $P_{\bf 1} (L_c)$ is exactly the plaquette projector $B_p$ used to define the Hamiltonian in Eq.~\eqref{eq:RSN} provided $L_c$ is the loop encircling (only) the plaquette $p$. 

In the case of a single plaquette, $\langle P_X(L_c) \rangle\neq 0$ iff $X\in \mathcal{F}$. Using the fact that $S$ is unitary and symmetric as well as Eq.~\eqref{eq:fluxonid}, one finds:
%
%
\be
\lim_{N_{\rm p}^{\mathcal{R}} \rightarrow \infty}  \langle P_{\bf 1}(L_c) \rangle =  \frac{{\rm e}^\beta}{{\cal D} - 1 + {\rm e}^\beta}=-e,
\label{eq:nrjproj}
\ee
%
%
where $e$ is given in Eq.~\eqref{eq:nrj} when $X=\mathbf{1}$, and
%
%
\be
\lim_{N_{\rm p}^{\mathcal{R}} \rightarrow \infty}  \langle P_X(L_c) \rangle =  \frac{d_X n_{X,1}}{{\cal D} - 1 + {\rm e}^\beta},
\ee
%
%
when $X\in \mathcal{F}^*$, i.e., is a nontrivial fluxon. This thermal average of a nontrivial fluxon projector has been computed in Ref.~\cite{Hu14} in the Fibonacci SN model and used to discuss an effective Pauli exclusion principle (see also Ref.~\cite{Vidal22}).

%
%
\section{Wegner-Wilson loops}
\label{sec:wwlconf}
%
%

%
%
\subsection{From projectors to WWL}
%
%

%
%
\begin{figure}[t]
\includegraphics[width=0.4\columnwidth]{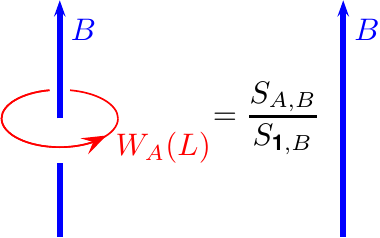}
\caption{A Wegner-Wilson loop operator $W_A$ acting on a string $B$.  This is known as the ``unlinking'' relation.}
\label{WWloop}
\end{figure}
%
%
Wegner-Wilson loops (WWL) are a different type of string operator, which should be thought of as quasiparticle world lines.  These are closely related to the projection operators $P_A(L)$ discussed in the previous section.  For any given closed loop $L$ on our surface, the unlinking relation  (see, e.g., Refs.~\cite{Kitaev06,Simon_book}) shown in Fig.~\ref{WWloop} directly gives:
%
%
\be
W_A(L)=\sum_{B \in {\cal Z}({\cal C})}  \frac{S_{A,B}}{S_{{\bf 1},B}} P_B(L).
\label{eq:WP}
\ee
%
%
Using the unitarity of $S$, one also gets the inverse transformation
%
%
\be
P_A(L)=S_{{\bf 1},A}\sum_{B  \in {\cal Z}({\cal C})} S_{A,B}^* W_B(L) .
\label{eq:PW}
\ee
%
%
Using Eqs.~\eqref{eq:Pidid} and \eqref{eq:WP}, it is easy to show that $W_{\bf 1}$ is actually the identity operator $\mathds{1}$.

The closed string operators $W_A$ satisfy the fusion algebra
%
%
\be
W_A(L) W_B(L) = \sum_{C  \in {\cal Z}({\cal C})} N_{A,B}^C W_C(L),
\label{eq:WW}
\ee 
%
%
where $N_{A,B}^C$ are the fusion multiplicity coefficients of ${\cal Z}({\cal C})$.  Equation~\eqref{eq:WW} can be proven by using the Verlinde formula \eqref{Wq2} along with the unitarity of $S$ and the projector property of $P_A(L)$ given in Eq.~\eqref{eq:PP0}.

A remarkable case of Eq.~(\ref{eq:PW}) is obtained for the projector onto the vacuum $A=\mathbf{1}$ for which  
%
%
\be
P_\mathbf{1}(L)=\sum_{ B \in  {\cal Z}({\cal C})} \frac{d_B}{\mathcal{D}^2} W_B(L). 
\ee
%
%
A slightly different normalization of this object,
%
%
\be
 \Omega(L) = {\cal D} P_\mathbf{1}(L), 
 \label{eq:OmegaDef}
\ee
%
%
is known  as a Kirby strand~\cite{Simon_book}, and is used extensively in Appendix~\ref{app:surg}. As explained in Sec.~\ref{sub:contra}, if $L$ surrounds a single plaquette $p$, then $P_\mathbf{1}(L)=B_p$.

Using the results obtained in Sec.~\ref{sec:proj} for the projectors, it is straightforward to obtain the thermal average of WWL operators using  \eqref{eq:WP}. For instance, for a large genus-$g$ surface, at any nonzero temperature, and for arbitrary couplings, one gets from  Eqs.~\eqref{eq:Mratg1} and \eqref{eq:WP}
%
%
\be
 \lim_{N_{\rm p} \rightarrow \infty} \langle W_A(L_h) \rangle = \frac{\sum_{C \in {\cal P}} N_{A C}^{C}  \,\, S_{\mathbf{1},C}^{2 - 2g}}{\sum_{C \in {\cal P}} S_{\mathbf{1},C}^{2 - 2g}},  
 \ee
%
%
where we used the Verlinde formula \eqref{Wq2}.  Such a WWL around a handle was studied, for example, in Ref.~\cite{Nussinov09_1} in a similar context. It shows a different behavior at zero temperature:
%
%
\be
 \lim_{T \to 0} \langle W_A(L_h) \rangle = \frac{\sum_{C \in {\cal Z(C)}} N_{A C}^{C}  \,\, S_{\mathbf{1},C}^{2 - 2g}}{\sum_{C \in {\cal Z(C)}} S_{\mathbf{1},C}^{2 - 2g}},  
 \ee
%
%
where we used Eqs.~\eqref{eq:PhandleSN} and \eqref{eq:WP}. Note how the Drinfeld center replaces the set of pure fluxons in the above sums when contrasting the $T=0$ and $T>0$ behaviors.

For the simple case of the SN model, one recovers the obvious fact that, at zero temperature, for any $A \in \mathcal{Z(C)}$, and for any contractible loop $L_c$,
%
%
\be
\lim_{T \rightarrow 0}\langle W_{A}(L_c) \rangle  = d_{A}.
\label{eq:WAT0}
\ee
%
%
Indeed, for any ground state of the SN model, one has only the vacuum in each plaquette. Thus, this result can be directly obtained from the unlinking relation (see Fig.~\ref{WWloop} for  $B={\bf 1}$). One can also obtain a simple expression of the WWL operators in the SN model at finite temperature, in the thermodynamic limit. Using 
Eqs.~\eqref{eq:50} and \eqref{eq:WP} one gets
%
%
\be
\lim_{N_{\rm p}^{\mathcal{R}} \rightarrow \infty}\langle W_A(L_c) \rangle = d_A \left(\frac{{\cal D} \frac{n_{A,1}}{d_A}-1+{\rm e}^{\beta}}{{\cal D} - 1 + e^\beta}   \right)^{N_{\rm p}^{\overline{\mathcal{R}}}}.
\label{eq:93}
\ee
%
%

%
%
\subsection{Confinement}
%
%
The behavior of $\langle W_A (L) \rangle$ when the size of the region $\mathcal{R}$ encircled by the loop $L$ varies can be used to diagnose the confinement and deconfinement properties of the quasiparticle $A$ (see, e.g., Ref.~\cite{Ritz21} for a study of the WWL operators in the SN model perturbed by a string tension~\cite{Schulz13,Schotte19}). Typically, in the case of a contractible loop $L_c$, a deconfined quasiparticle $A$ corresponds to a ``perimeter law,''  i.e.,  the WWL decays exponentially with the perimeter $|L_c|$ of the loop $L_c$
%
%
\be
\langle W_A(L_c) \rangle \sim {\rm e}^{- c_1 |L_c|},
\ee
%
%
when $L_c$ grows and where $c_1>0$. In the extreme case where $c_1=0$ (no decay of the WWL), it is known as the ``zero law"\cite{Hastings05}. 

A confined quasiparticle $A$ corresponds to an ``area law,''  i.e.,  the WWL decays exponentially with  the area $\mathcal{A}$ of the region delimited by $L_c$ (i.e., $\mathcal{R}$ or $\overline{\mathcal{R}}$ since we are on a closed surface)
%
%
\be
\langle W_A(L_c) \rangle \sim {\rm e}^{- c_2 \mathcal{A}},
\ee
%
%
when $L_c$ grows and where $c_2>0$.

Let us analyze the confining properties of the quasiparticles in the special case of the SN model. As can already be seen in Eq.~\eqref{eq:WAT0},  at zero temperature,  $W_A(L_c)$ obeys a  zero law (see Ref.~\cite{Ritz21} for further discussions). This means that all quasiparticles are deconfined in the ground state, which is a fingerprint of a topologically ordered phase. 
By contrast, at any finite temperature, Eq.~\eqref{eq:93} indicates an area law, i.e., a confinement, for all particles of the Drinfeld center except for pure fluxons identified by \mbox{$n_{A,1}=d_A$}. Indeed, for pure fluxons, one always has
%
%
\be
\langle W_A(L_c)\rangle = d_{A}, 
\label{eq:puref}
\ee
%
%
indicating a deconfinement of these particles at any temperature.  Equation \eqref{eq:puref} suggests that pure fluxons are insensitive to the presence of other fluxons in the system, i.e., that they braid trivially with all fluxons.  This trivial braiding property,  already mentioned in Sec.~\ref{sub:ZRSN}, can be summarized as follows: {\it if $A$ is a pure fluxon, i.e, if $n_{A,1}=d_A$, then one has $S_{A,B}=d_A d_B /\mathcal{D}$, for any fluxon}. As a byproduct, this property also holds for any particle $B \in \mathcal{F}^\otimes$.
For all other particles $A\notin \cal{F}$, one has $n_{A,1} < d_{A}$, and Eq.~\eqref{eq:93} describes an area law. More generally, one may write 
%
%
\be
\langle W_A(L_c) \rangle = d_A \,\, {\rm e}^{-N_{\rm p}^{\overline{\mathcal{R}}}/N_A^*},
\label{eq:63}
\ee
%
%
where we introduced the temperature-dependent characteristic area 
%
%
\be
 N_A^* = \left[\ln\left( \frac{{\cal D}  -1 + {\rm e}^{\beta}}{{\cal D} \frac{n_{A,1}}{d_A} -1 + {\rm e}^{\beta}}  \right)\right]^{-1}, 
 \label{eq:64}
 \ee
%
%
which diverges for $A\in {\cal P}\:  (n_{A,1}= d_A)$  and is minimum for $A \notin {\cal F} \: (n_{A,1}=0)$. In other words, particles that are not fluxons are strongly confined.

%
%
\section{Topological mutual information}
\label{sec:tmi}
%
%
One way to characterize topological order at zero temperature is to compute the topological entanglement entropy introduced in Refs.~\cite{Kitaev06b,Levin06} (see also Ref.~\cite{Hamma05}). However, at finite temperature,  this quantity, defined as the constant term of the von Neumann entanglement entropy, suffers from several problems which have led Iblisdir {\it et al.} to rather consider the topological mutual information $I_{\rm topo}$~\cite{Iblisdir09,Iblisdir10}. The main issue is that the finite-temperature entanglement entropy no longer follows an area law but also features an extensive term. It is therefore no longer symmetric between the inner and outer regions of the contour, and cannot be thought of as measuring only the entanglement between these regions. 

%
%
\subsection{Definitions and conjecture}
\label{sec:conjec}
%
%
For a bipartition of the system into two regions $\mathcal{R}$ and $\overline{\mathcal{R}}$, the mutual information is defined as:
%
%
\be
I_\mathcal{R}=S_\mathcal{R}+S_{\overline{\mathcal{R}}}-S_{\mathcal{R} \cup \overline{\mathcal{R}}}=I_{\overline{\mathcal{R}}},
\label{eq:Idef}
\ee
%
%
where $S_\mathcal{R}=-\tr_\mathcal{R} \: \rho_\mathcal{R} \ln  \rho_\mathcal{R}$ is the von Neumann entropy, and $\rho_\mathcal{R}=\tr_{\overline{\mathcal{R}}} \: ({\rm e}^{-\beta H})/Z$. In the limit where the  length $|L_c|$ of the boundary $L_c$ between the two regions goes to infinity, one expects the mutual information to behave as~\cite{Iblisdir09,Iblisdir10} 
%
%
\be
I_\mathcal{R}=\alpha |L_c| -\gamma.
\label{eq:Igamma}
\ee
%
This is  often called an ``area law" although it implies the perimeter of the loop $L_c$.
The topological mutual information is then defined as $I_{\rm topo}=-\gamma$. Under some simple assumptions, Iblisdir {\it et al.} conjectured a general form of $I_{\rm topo}$ at finite temperature for the Kitaev quantum double model~\cite{Kitaev03} based on the Kullback-Leibler divergence~\cite{Iblisdir09,Iblisdir10}. For a surface with $g=0$, and in the limit where $|L_c| \to \infty$, Iblisdir {\it et al.} conjectured that
%
%
\be
I_{\rm topo} (T)=- \sum_{A \in {\cal Z}({\cal C})} \langle P_A(L_c) \rangle \ln \left[\langle P_A(L_c) \rangle\frac{\mathcal{D}^2 }{d_A^2}\right],
\label{eq:Itopo}
\ee
%
%
where $P_A(L_c)$ is the projector whose thermal average is given in Eq.~\eqref{eq:Pcont}. As explained in Sec.~\ref{sub:contra}, it is nonvanishing only if $A \in {\cal F}^{\otimes}$.

Although the conjecture \eqref{eq:Itopo} has been derived in a different context, it is based on general assumptions~\cite{Iblisdir09,Iblisdir10}. In the following, we will assume that it holds for the SN model on which we focus. As we shall see, it reproduces the exact results in the zero-$T$ and infinite-$T$ limits. 
Proving this conjecture for arbitrary temperature is work in progress.

%
%
\subsection{Infinite-temperature limit}
\label{sec:highT}
%
%
Generically, one expects zero mutual information at infinite temperature due to a loss of quantum correlations.  However, for the SN model, some nontrivial residual information remains in our system because we are working in a restricted Hilbert space (vertex constraint)~\cite{Hermanns14}.

The general expression of $\langle P_A(L_c) \rangle$ is given in Eq.~\eqref{eq:Pcont}. Here, for simplicity, we consider the thermodynamic limit where both $N_{\rm p}^{\mathcal{R}}$ and $N_{\rm p}^{\overline{\mathcal{R}}}$ go to infinity which, using Eq.~\eqref{eq:50}, yields
%
%
\be
\lim_{N_{\rm p}^{\mathcal{R}}, N_{\rm p}^{\overline{\mathcal{R}}}\rightarrow \infty}  \langle P_A(L_c) \rangle = d_A \sum_{C \in \cal{P}} S_{\bar A,C} S_{{\bf 1},C}= \frac{d_A^2}{{\cal D}^2} M_0,
\label{eq:68}
\ee
%
%
for $A \in {\cal F}^\otimes$, and $0$ otherwise [$M_0$ is defined in Eq.~\eqref{M}]. This is valid at any finite temperature (but not at $T=0$). Using Eq.~\eqref{eq:Pidid}, one then obtains
%
%
\be
I_{\rm topo} (T=\infty)=-\ln M_0.
\label{Itopo_infty}
\ee
%
%
As mentioned above, this nonvanishing contribution only stems from the fact that, for the model at hand, we work in a restricted Hilbert space. It has no quantum origin, and it does not reflect any long-range entanglement feature of the system. However, it does reflect some topological property, as only vertex configurations resulting from the fusion rules are allowed.

The result \eqref{Itopo_infty} is compatible with the one given in Ref.~\cite{Hermanns14} for the topological ``classical" entropy. Indeed, at infinite temperature, it was shown in that work, for some special input categories $\mathcal{C}$, that the topological classical entropy is given by
%
%
\be
 S_{\mathcal{R},{\rm topo}}(T=\infty) = -(m_{\overline{\mathcal{R}}}-1) \ln M,
 \label{eq:Stopo2}
\ee
%
%
where $m_{\overline{\mathcal{R}}}$ is the number of disconnected pieces in $\overline{\mathcal{R}}$, and where $M$ was identified as the number of Abelian particles in the input category $\mathcal{C}$~\cite{Hermanns14}. For the case, $m_{\mathcal{R}}=m_{\overline{\mathcal{R}}}=1$, considered here, this expression of the entropy gives $I_{\rm topo} (T=\infty)=-S_{\mathcal{R} \cup \overline{\mathcal{R}},{\rm topo}}=-\ln M$ (see Ref.~\cite{Hermanns14}). Interestingly, this expression is similar to Eq.~\eqref{Itopo_infty} and in agreement with our expression of the constant (topological) term in Eq.~\eqref{eq:enttot} provided $M=M_0$. It turns out that the input categories considered in Ref.~\cite{Hermanns14}  are  either Abelian or modular. In both cases, it is easy to prove that $M_0=M$. However, in general,  $M_0$ and $M$ may be different. For instance, if $\mathcal{C}={\rm Rep}(S_3)$, one has $M=2$ (since the group $S_3$ has two one-dimensional irreducible representations), but $M_0=1$ (since the only pure fluxon in $\mathcal{Z}[{\rm Rep}(S_3)]$ is the vacuum~\cite{Ritz24}). 

%
%
\subsection{Zero-temperature limit }
%
%
In the zero-$T$ limit, every plaquette is in the vacuum ($A={\bf 1}$) state. Thus, $\langle P_A(L_c) \rangle$ is given by Eq.~\eqref{eq:zeroTPSN} and one readily gets from Eq.~\eqref{eq:Itopo}:
%
%
\be
I_{\rm topo} (T=0) = -2\ln \mathcal{D},
\label{Itopo_zero}
\ee
%
%
which is the well-known zero-temperature result for a topological phase with total quantum dimension \mbox{$\mathcal{D}$~\cite{Kitaev06b,Levin06,Iblisdir10}}. 
%
%
\begin{figure}[t]
\includegraphics[width=0.9\columnwidth]{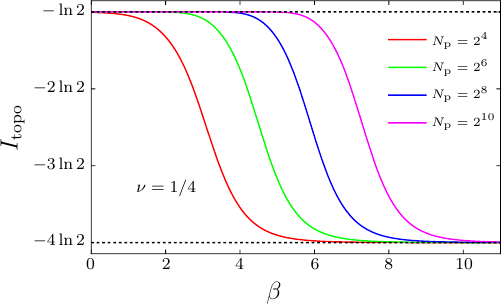}
\caption{Topological mutual information of  SN model for the Ising category ($\mathcal{D}=4$, $M_0=2$) as a function of  $\beta=1/T$ at fixed $\nu=1/4$. \mbox{$I_{\rm topo}(T=0)=-2\ln \mathcal{D}$} and \mbox{$I_{\rm topo}(T=\infty)=-\ln M_0$} are indicated by dashed lines.
}
\label{fig:Itopo_Ising_v0}
\end{figure}
%
%

%
%
\subsection{Finite temperature and scaling behavior}
%
%
Away from the two extreme cases discussed above, the situation is more subtle.  As can already be inferred from Eq.~\eqref{eq:50}, using the same arguments as in the previous section, one always gets
%
%
\be
 \lim_{N_{\rm p}^{\mathcal{R}}, N_{\rm p}^{\overline{\mathcal{R}}}  \rightarrow \infty}   I_{\rm topo}(T > 0)=-\ln M_0, 
 \label{eq:Stopo}
\ee
%
%
which is also the infinite-temperature limit [see Eq.~\eqref{Itopo_infty}].  This indicates that topological quantum order is destroyed in the thermodynamic limit for any $T>0$ as anticipated by Hastings~\cite{Hastings11}. This phenomenon is similar to the one observed in the toric code~\cite{Kitaev03} (see Refs.~\cite{Castelnovo07_2,Nussinov08,Iblisdir09,Iblisdir10}). 

For a given contour $L_c$, one observes a crossover between a low-$T$ region where $I_{\rm topo} \simeq -2\ln \mathcal{D}$, and a high-$T$ region where $I_{\rm topo} \simeq -\ln M_0$ (see Fig.~\ref{fig:Itopo_Ising_v0}). When increasing the system size (while keeping $L_c$ fixed), $I_{\rm topo}$ converge towards a unique curve (not shown). The crossover temperature can be estimated from Eq.~\eqref{eq:50} as follows. 

The dominant behavior of $\langle P_X(L_c) \rangle$ in the thermodynamic limit comes from pure fluxons and is given in Eq.~\eqref{eq:68}. The first finite-size correction comes from the nonpure fluxon (call it $C$) with largest ratio \mbox{$n_{C,1}/d_C<1$}:
%
%
\be
\langle P_X(L_c) \rangle -\frac{d_X^2}{\mathcal{D}^2}M_0 \simeq  \frac{d_X d_C}{\mathcal{D}} S_{\bar X,C} \, \text{e}^{-N_{\rm p}^{\overline{\mathcal{R}}}/N_\text{p}^*},
\ee
%
%
with the characteristic area 
%
%
\be
 N_\text{p}^* = \underset{A\notin \mathcal{P}}{\text{max}} \, N_A^* =\left[\ln\left( \frac{{\cal D}  -1 + {\rm e}^{\beta}}{{\cal D} \frac{n_{C,1}}{d_C} -1 + {\rm e}^{\beta}}  \right)\right]^{-1},
 \label{eq:Npstar}
\ee
%
%
where $N_A^*$ is given in Eq.~\eqref{eq:64}. The quantity $N_\text{p}^*$ reaches a constant $1/\ln \frac{d_C}{n_{C,1}}$ in the high-$T$ limit and diverges as
%
%
\be
 N_\text{p}^* \simeq \frac{\text{e}^\beta}{\mathcal{D}(1-\frac{n_{C,1}}{d_C})}
\ee
%
%
in the low-$T$ limit. 
The crossover temperature is reached when $N_\text{p}^*\simeq N_{\rm p}^{\overline{\mathcal{R}}}$ and is roughly given by 
%
%
\beqn
T_c \simeq \frac{1}{\ln [N_{\rm p}^{\overline{\mathcal{R}}} \mathcal{D}(1-n_{C,1}/d_C)]} \simeq \frac{1}{\ln N_{\rm p}^{\overline{\mathcal{R}}}}
\eeqn
%
%
when $N_{\rm p}^{\overline{\mathcal{R}}}$ is large. Such a behavior is analogous to that of the 1D classical Ising model, which has a vanishing critical temperature. That the toric code model is in the same universality class as the 1D classical Ising model is well-known (see, e.g. Ref.~\cite{Weinstein19}). Here, we suspect that this is also the case of the SN model for any input category.

%
%
\begin{figure}[t]
\includegraphics[width=0.9\columnwidth]{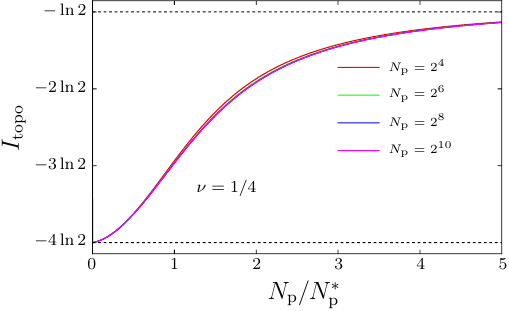}
\caption{Topological mutual information of the Ising SN model as a function of the scaling variable $N_{\rm p}/N_{\rm p}^*$ at fixed $\nu=1/4$, where $N_{\rm p}^*$ is defined in Eq.~\eqref{eq:Npstar}.}
\label{fig:Itopo_Ising_v1}
\end{figure}
%
%

A close inspection  shows a nontrivial interplay between the temperature $T$ and the total
system size \mbox{$N_{\rm p}=N_{\rm p}^{\mathcal{R}}+N_{\rm p}^{\overline{\mathcal{R}}}$}, similar to the one found in Refs.~\cite{Castelnovo07_2,Iblisdir09,Iblisdir10} for the toric code. More precisely, if one sets $N_{\rm p}^{\overline{\mathcal{R}}}=\nu N_{\rm p}$, with a fixed ratio, $0 < \nu < 1 $, one can show that, in the large-$N_{\rm p}$ limit, $I_{\rm topo}$ depends on $\nu$ and on the proper scaling variable $N_{\rm p}/N_{\rm p}^*$,
%
%
%
%
where $N_p^*$ is defined in Eq.~\eqref{eq:Npstar}. 

Such a scaling law is similar to the one observed in Refs.~\cite{Castelnovo07_2,Iblisdir09,Iblisdir10} for the toric code and indicates that topological quantum order can persist at finite temperature provided the system size is small enough. Differences between the Kitaev quantum double and the SN models will be discussed elsewhere.

We display in Fig.~\ref{fig:Itopo_Ising_v0} the topological mutual information $I_{\rm topo}$ as a function of the inverse temperature for various system sizes at ratio $\nu=1/4$. One clearly observes that when the system size $N_{\rm p}$ increases, the topological order characterized by $I_{\rm topo}=-4\ln 2$ is destroyed at a temperature which decreases and vanishes in the thermodynamic limit. However, when plotted as a function of the scaling variable $N_{\rm p}^*$ (see Fig.~\ref{fig:Itopo_Ising_v1}), $I_{\rm topo}$ converges towards a ``universal" function which interpolates between $-4 \ln 2$ at $T=0$ [see Eq.~\eqref{Itopo_zero}] and $-\ln 2$ at $T=\infty$ [see Eq.~\eqref{Itopo_infty}]. 

In summary, the crossover line $N_{\rm p}^*=N_{\rm p}^{\overline{\mathcal{R}}}$ separates two domains in the $(T,N_{\rm p}^{\overline{\mathcal{R}}})$ plane: a low-$T$ and small-$N_{\rm p}^{\overline{\mathcal{R}}}$ domain with $I_{\rm topo} \simeq -2\ln \mathcal{D}$ (indicative of topological order),  and a high-$T$ and large-$N_{\rm p}^{\overline{\mathcal{R}}}$ domain with \mbox{$I_{\rm topo} \simeq -\ln M_0$} (absence of topological order).

%
%
\section{Conclusions and perspectives}
\label{sec:ccl}
%
%
In this work, we studied the finite-temperature properties of RSN model which is an extension of the original SN model. Using an exact expression of the degeneracies obtained in a companion paper (see Ref.~\cite{Ritz24}), we provided a simple and exact expression of the partition function [see Eq.~\eqref{eq:ZRSN}] which is valid for any UFC, any trivalent graph, and any orientable closed surface. This partition function allowed us to analyze the finite-temperature behavior of several quantities. In particular, using simple surgery arguments, we computed the thermal average of the projector onto a given particle sector for three different types of closed loop: handle, throat, and contractible [see Eq.~\eqref{eq:exphand} and Fig.~\ref{fig:fidget}, Eq.~\eqref{eq:expth} and Fig.~\ref{fig:sewg1g2}, and Eq.~\eqref{eq:Pcont}, respectively]. These projectors are directly related to WWL operators [see Eq.~\eqref{eq:WP}] which provide some information about the confinement of the excitations. Interestingly, these projectors are the key ingredients to compute the topological mutual information according to a conjecture [see Eq.~\eqref{eq:Itopo}] proposed by Iblisdir {\it et al.}~\cite{Iblisdir09,Iblisdir10}.

Our main finding is that, as seen from different probes (such as the specific heat,  confinement, or entanglement), topological order does not survive at $T>0$ in the thermodynamic limit. This is in line with exact results~\cite{Hastings11}. However, we find that there is a scaling behavior between the temperature and the system size, generalizing what was found by Iblisdir et al.~\cite{Castelnovo07_2,Nussinov08,Iblisdir09,Iblisdir10}. This is essentially the same scaling as that of the 1D classical Ising model.

Importantly, we have identified different subsets of objects in the Drinfeld center $\mathcal{Z(C)}$ that play a special role:  pure fluxons $\mathcal{P}$ that drive the thermodynamic limit and braid trivially with all fluxons, the fluxons $\mathcal{F}$ that characterizes plaquette excitations, and the fusion product of fluxons $\mathcal{F}^\otimes$ (see Appendix~\ref{app:transpflu} for more details). The content of these subsets not only depends on the Drinfeld center, but also on the input category. In other words,  two categories with the same Drinfeld center (i.e., Morita equivalent) may have different such subsets. 

We have also shown that the Drinfeld center ${\cal Z}({\cal C})$ is enough to understand the $T=0$ or ground-state properties, but that it is not sufficient to analyze properties at finite temperature. For the latter, one needs to know more information that depends on the input category $\cal C$. More precisely, one needs to determine the internal multiplicities $n_{A,1}$ from the tube algebra.    Perhaps this is not surprising given our prior work in Ref.~\cite{Ritz24} where we showed that the nontopological degeneracies of the string net models depend on  $n_{A,1}$. 

For future work we plan on studying models featuring all types of excitations in the Drinfeld center (not only fluxons). This could be done, for instance, in the Kitaev quantum double model~\cite{Kitaev03} or in the extended Levin-Wen model~\cite{Hu18}, both allowing vertex excitations. Such a study will allow us to analyze the interplay between the various excitations of topologically ordered phases. 

\acknowledgments We thank N. Bultinck, B. Dou\c{c}ot, S. Iblisdir and L. Lootens, for fruitful discussions. We acknowledge the financial support of the Emergence programs iMAT and TQCNAA at Sorbonne Universit\'e, Paris.

\appendix

%
%
\section{Fluxons, pure fluxons and fusion product of fluxons}
\label{app:transpflu}
%
%
Simple objects $A$ of the Drinfeld center $\mathcal{Z}(C)$ index anyon types, among which we can still distinguish subtypes $(A,s)$ which are, in addition, indexed by an object $s\in \mathcal{C}$~\cite{Hu18}. These subtypes also correspond to ``diagonal sectors'' of the tube algebra (see Appendix~\ref{app:fusiontubes}). Fluxons (or plaquette excitations) are identified as those subtypes which are indexed by the identity of $\mathcal{C}$, $(A,1)$. The multiplicities $n_{A,s}$ count the number of subtypes $(A,s)$ inside an anyon $A$. Therefore, another way to define fluxons (elements of $\mathcal{F}$) is
%
%
\be
 A\in \mathcal{F} \Leftrightarrow n_{A,1}>0.
\ee
%
%
Among fluxons, we distinguish ``pure fluxons'' (elements of $\mathcal{P}$)~\cite{Hu18} defined as 
%
%
\be
A\in \mathcal{P} \Leftrightarrow n_{A,1}=d_A  .
\label{eq:neqd}
\ee
%
%
Since $d_A=\sum_s n_{A,s} d_s$, where $d_s$ are the quantum dimensions of the simple objects in $\mathcal{C}$ and $d_A$ the quantum dimensions of the anyons $A \in \mathcal{Z(C)}$, this means that a pure fluxon contains only the subtype corresponding to the trivial input object $n_{A,s}=d_A \delta_{s,1}$, hence its name.

Below, we prove that this definition \eqref{eq:neqd} is equivalent to the following: 
%
%
\be
A\in \mathcal{P} \Leftrightarrow S_{A,B}=\frac{d_A d_B}{\mathcal{D}}, \, \forall B \in \mathcal{F}.
\label{eq:SddD}
\ee
%
%
This means that the braiding between a pure fluxon and a fluxon is trivial so that one may unknot the $A$ and $B$ loops in the definition of the $S$-matrix leading to Eq.~(\ref{eq:SddD}) (see, e.g. Fig.~6 in~\cite{Ritz24}). In other words, a pure fluxon is transparent to all fluxons (but not necessarily to nonfluxons).  A corollary (from the hexagon equation~\cite{Simon_book,Kitaev06}) is that Eq.~\eqref{eq:SddD} also holds if $A$ is a pure fluxon and $B$ is any fusion product of fluxons ($B \in {\cal F}^{\otimes}$). Obviously, the three sets $\mathcal{P}$, $\mathcal{F}$ and $\mathcal{F}^\otimes$ are such that:
%
\be
\{\mathbf{1}\} \subseteq  \mathcal{P} \subseteq \mathcal{F} \subseteq \mathcal{F}^\otimes \subseteq \mathcal{Z(C)}.
\ee
%
%

From Eqs.~\eqref{eq:Pidid} and \eqref{eq:68} one obtains the following equality
%
%
\be
\sum_{A\in \mathcal{Z(C)}}d_A^2 = \sum_{B\in \mathcal{P}}d_B^2 \times \sum_{C\in \mathcal{F}^\otimes}d_C^2,
\ee
%
%
and the corresponding inequality
%
%
\be
\mathcal{D}^2\geq M_0 \, |\mathcal{F}^\otimes|,
\ee
%
%
which  relates the number of quasiparticles that can be obtained by fusion of fluxons, $|\mathcal{F}^\otimes|$, to the total quantum dimension $\mathcal{D}$ and to $M_0$ (which, in the case of a commutative input category, is the number of pure fluxons). It shows that when $M_0$ is small $|\mathcal{F}^\otimes|$ is large and vice-versa.

\subsection{Pure fluxons braid trivially with fluxons}
We now prove that (\ref{eq:SddD}) and (\ref{eq:neqd}) are equivalent.

It is easy to prove that the right-hand side of Eq.~(\ref{eq:SddD}) implies the right-hand side of Eq.~(\ref{eq:neqd}). We start from the fluxon identity \eqref{eq:fluxonid}~\cite{Ritz24}
%
%
\be
n_{A,1}=\sum_{B  \in {\cal Z}({\cal C})} S_{A,B}\, n_{B,1} = \sum_{B \in \mathcal{F}} S_{A,B}\,  n_{B,1},
\label{eq:A5}
\ee
%
%
and note that, on the right-hand side, only fluxons play a role. Therefore, if $A$ is a pure fluxon, we can use Eq.~(\ref{eq:SddD}) in the right-hand side to get
%
%
\be
n_{A,1}=\sum_{B \in \mathcal{F}} \frac{d_A d_B}{\mathcal{D}} n_{B,1}= d_A \:\: \frac{1}{\mathcal{D}} \sum_{B \in \mathcal{F}} d_B n_{B,1}.
\ee
%
%
Then, we can use Eq.~\eqref{eq:A5} again for the special case $A={\bf 1}$ ($n_{{\bf 1},1} = 1$), to prove that: 
%
%
\be
\frac{1}{\mathcal{D}} \sum_{B \in \mathcal{F}} d_B n_{B,1}=1,
\label{eq:identity1}
\ee
%
%
and we obtain Eq.~(\ref{eq:neqd}).  More generally,  one has~\cite{Neupert16}
%
%
\be
\frac{1}{\mathcal{D}} \sum_{B  \in {\cal Z}({\cal C})} d_B n_{B,s}= d_s.
\ee
%
%
for all $s \in \mathcal{C}$.

In order to prove that the right-hand side of Eq.~(\ref{eq:neqd}) implies the right-hand side of Eq.~(\ref{eq:SddD}), we start from the following expression for the $S$-matrix (see, e.g., Eq.~(223) in Ref.~\cite{Kitaev06})
%
%
\be
  S_{A,B}=\frac{1}{\mathcal{D}} \sum_{C  \in {\cal Z}({\cal C})} N_{A,\bar{B}}^C \frac{\theta_C}{\theta_A \theta_B} d_C,
  \label{eq:SABform}
\ee
%
%
which we apply to the case where $A\in \mathcal{P}$ and  $B \in \mathcal{F}$. As fluxons, $A$ and $B$ have trivial twists ($\theta_A=\theta_B=1$). Indeed, the vector  ${\bf n}_1$  with components $n_{C,1}$ (\mbox{$n_{C,1}=0$}, if $C\notin  {\cal F}$)~\cite{Ritz24}, also obeys $T.{\bf n}_1={\bf n}_1$, where the twist $T-$matrix is defined by $T_{AB}=\theta_A\delta_{AB}$ (see, e.,g., Ref.~\cite{Ritz24}). Thus, Eq.~\eqref{eq:SABform} gives

%
%
\be
S_{A,B}=\frac{1}{\mathcal{D}} \sum_{C  \in {\cal Z}({\cal C})} N_{A,\bar{B}}^C \theta_C d_C .
\ee
%
%
Then, we use the fact that the fusion of a pure fluxon $A$ with a fluxon $B$ can only be a fluxon (see below) so that $\theta_C=1$ and one further gets
%
%
\be
S_{A,B}=\frac{1}{\mathcal{D}} \sum_{C  \in {\cal Z}({\cal C})} N_{A,\bar{B}}^C d_C =\frac{d_A d_B}{\mathcal{D}} ,
\ee
%
%
which completes the proof.

\subsection{Fusion of pure fluxon with fluxon gives fluxon}
In order to prove that the fusion product of a pure fluxon and a fluxon can only be a fluxon, we use the fusion of tubes as described in detail in Appendix~\ref{app:fusiontubes}. The crux of the argument is the following.  We consider the fusion of two anyon subtypes $(A,r)$ and $(B,s)$ that results in some  $(C,t)$, where $A,B,C$ are in $\mathcal{Z(C)}$ and $r,s,t$ are in $\mathcal{C}$.  If $A$ is a pure fluxon then it only contains the trivial input string, i.e. $r=1$. The fusion of $(A,1)$ with $(B,s)$ necessarily gives $(A\times B,1\times s)=(A\times B,s)$ so that $C \in A\times B$ and $t=s$. If $B$ is a fluxon ($n_{B,1}>0$), it means that it contains, at least,  $s=1$. Hence, the particle $C$ also has this sector $t=s=1$, which means that $C$ is also a fluxon.

As a corollary, we see that if $A$ and $B$ are both pure fluxons (i.e., if $r=1$ and $s=1$ are the only nonvanishing sectors), then $C$ must be a pure fluxon as well. In other words, pure fluxons are closed under fusion and therefore form a subcategory. This fusion category is also unitary and symmetric (and not modular). As all fluxons are bosons (i.e. have a trivial twist), we can use the known result that a symmetric fusion category made of bosons is equivalent to a Rep$(G)$ category (see, e.g.,~\cite{Simon_book}).

%
%
\section{Fusion of tubes}
\label{app:fusiontubes}
%
%

In this Appendix we compute the fusion outcome $C$ of two anyons $A$ and $B$  using the tube algebra. We also show that if $A$ is a pure fluxon and $B$ is a fluxon, then necessarily $C$ is also a fluxon (as anticipated above). 

%
%
\subsubsection{Fusion rules from the tube algebra}
%
%
One way to obtain the fusion rules from the tube algebra is to use the half-braidings to compute the $S-$matrix and to use the Verlinde formula to obtain the $N-$matrices. This is the way followed, e.g., in Refs~\cite{Levin05, Lan14, Ritz24}. Here, we follow a different path and obtain the fusion coefficients for the Drinfeld center directly from the tube algebra, without using the half-braidings.

We follow the general strategy described in Sec.~V in Ref.~\cite{Bultinck17} (see also Sec.~IV in Ref.~\cite{Delcamp17}). For simplicity, we concentrate on commutative and self-dual input categories, but the method can easily be generalized to the non-self-dual and noncommutative case. The first one only requires to put arrows on all strings, where reversing an arrow means going from an object $i \in \mathcal{C}$ to its dual. The second one implies to label sectors of the tube algebra by two indices $r, \alpha$ instead of one $r$, and to consider particles with $n_{A,1}\geqslant 1$ (see Appendix A in Ref.~\cite{Ritz24} for more details). In the following, by convention, we will denote simple objects in $\mathcal{C}$ by lowercase letters and simple objects in $\mathcal{Z}(\mathcal{C})$ by capital letters.

The (central) projector $P_A$ on a particle $A$ can be written as 
%
%
\be
P_A =\sum_r p_A^{rr}.
\ee
%
%
The simple idempotents $p_A^{rr}$ and nilpotents $p_A^{rs}$ decomposes in the tube basis as
%
%
\be
p_A^{rs}= \sum_{i, j\in \mathcal{C}}M^{-1}_{A,irsj} Q_{irsj},
\label{proj-tub}
\ee
%
%
where the coefficients $M^{-1}_A$ depending on $A \in \mathcal{Z}(\mathcal{C})$, $r,s \in \mathcal{C}$ label the sectors in the tube algebra, and where the $Q$'s are the tubes. The reverse version of this formula is
%
%
\be
Q_{irsj}=\sum_A M_{A, irsj}p_A^{rs}.
\label{tube-proj}
\ee
%
%
The $M$ and $M^{-1}$ together verify
%
%
\be 
\sum_{i,j}M^{-1}_{A,irsj}M_{B, irsj} = \delta_{A,B}.
\label{generalid}
\ee
%
%

\begin{figure}[h]
\includegraphics[width=0.2 \columnwidth]{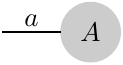}
\caption{A single quasiparticle basis state $\ket{A,a}$ corresponding to a simple idempotent $p_A^{aa}=|A,a\rangle \langle A,a|$, with $A\in \mathcal{Z(C)}$ and $a\in \mathcal{C}$.}
\label{basisstate}
\end{figure}

Following Ref.~\cite{Lan14}, we represent a one-quasiparticle basis state $\ket{A, a}$, with $A\in \mathcal{Z(C)}$ and $a\in \mathcal{C}$, as in Fig.~\ref{basisstate}.
%
%
%
%
The dimension of the corresponding Hilbert space is:
%
%
\be 
\label{eq:1qp}
\dim \mathcal{H}_{1\textnormal{QP}} = \sum_{A,a}  n_{A,a}.
\ee
%
%

\begin{figure}[h]
\includegraphics[width=0.3 \columnwidth]{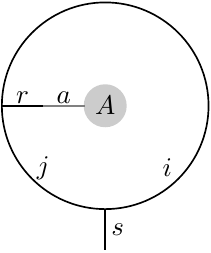}
\caption{Action of a tube on a one-quasiparticle state $\ket{A,a}$.}
\label{fig:projector}
\end{figure}

The action of a tube on this basis state, as represented on Fig.~\ref{fig:projector}, is 
%
%
\be
Q_{irsj} \ket{A, a} = \delta_{r, a} M_{A, irsj} \ket{A,s}.
\label{tubeaction}
\ee
%
%

%
%
%
%

\begin{figure}[h]
\includegraphics[width=0.44\columnwidth]{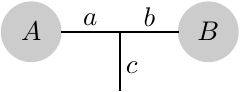}
\caption{A two-quasiparticle state $\ket{A, a, B, b, c}$.}
\label{2qp}
\end{figure}

We now introduce a basis for a two-quasiparticle state. We write a state in this basis $\ket{A, a, B, b, c}$ and represent it graphically as in Fig.~\ref{2qp}. 
%
%
%
%

\begin{figure}[h]
\includegraphics[width=0.45 \columnwidth]{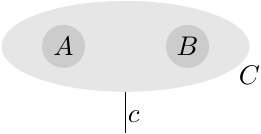}
\caption{A two-quasiparticle state $\ket{A, B, C, c}$.}
\label{2qpbis}
\end{figure}

In other words, a two-quasiparticle state is entirely fixed if we specify the two quasiparticle types $A$ and $B$, their tube-algebra sectors $a$ and $b$, and the channel $c$ in which $a$ and $b$ fuse. An alternative way to describe the two-quasiparticle Hilbert space is to take states $\ket{A, B, C, c}$ as a basis (see Fig.~\ref{2qpbis}). 
%
%
%
%
Here, we fix $A$, $B$ and their fusion outcome $C$ as well as the tube algebra sector $c$ of $C$. These states are eigenvectors of the simple idempotents $p_C^{cc}$: 
%
%
\be
p_{C'}^{c'c'}\ket{A, B, C, c}=\delta_{c,c'}\delta_{C,C'}\ket{A, B, C, c}.
\ee
%
%
The dimension of the two-quasiparticle Hilbert space is
%
%
\beqn 
\label{eq:2qp}
\dim \mathcal{H}_{2\textnormal{QP}} &=& \sum_{A, B, a, b, c}N_{a,b}^cn_{A,a}n_{B,b},\\
&=& \sum_{A, B, C, c}N_{A, B}^C n_{C,c}.
\eeqn
%
%
 The equality between these two lines follows from the commutation between fusion and restriction in anyon condensation~\cite{Neupert16}. Since
%
%
\be
\sum_{A, B, a, b}n_{A,a}n_{B,b} \sum_c N_{a,b}^c \geq \sum_{A, B, a, b}n_{A,a}n_{B,b},
\ee
%
%
the dimension of the two-quasiparticle Hilbert space \eqref{eq:2qp} is larger than the square of the dimension of the one-quasiparticle Hilbert space \eqref{eq:1qp}, which is a signature of entanglement.

\begin{figure}[h]
\includegraphics[width=0.41 \columnwidth]{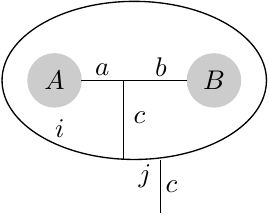}
\caption{Tube $Q_{iccj}$ acting on $\ket{A, a, B, b, c}$.}
\label{fig:projbig}
\end{figure}

\begin{figure}[h]
\includegraphics[width=0.55 \columnwidth]{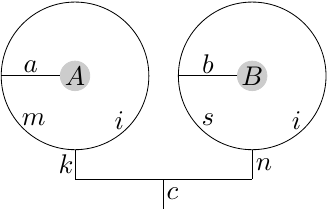}
\caption{Same as Fig.~\ref{fig:projbig} after a series of $F$-moves. Now, there are two tubes $Q_{iakm}$ and $Q_{ibns}$ acting on single quasiparticle states $|A,a\rangle$ and $|B,b\rangle$.}
\label{fig:proj3}
\end{figure}

The action of a tube $Q_{iccj}$ on a state $\ket{A,a, B,b, c}$ can graphically be represented as in Fig.~\ref{fig:projbig}. 
%
%
%
%
Applying a series of $F$-moves, this diagram can be modified into the diagram of Fig.~\ref{fig:proj3}.
%
%
%
%
In this figure, the diagram of the big tube around $A$ and $B$ has been reduced to two smaller tubes, one tube $Q_{iakm}$ around $A$ and $Q_{ibns}$ around $B$. Using finally Eq.~\eqref{tubeaction} we see that we can write 
%
%
\begin{align}\nonumber Q_{iccj}&\ket{A,a,B,b, c} = \!\!\!\!\! \sum\limits_{\substack{
i,j, s,r,\\ k, m, n}}F^{ii0}_{aar}F^{ii0}_{bbs}F^{sib}_{air}F^{src}_{ijk}F^{skj}_{cin}F^{iar}_{ikm}\\ &\times\sqrt{\frac{d_i d_a}{d_r}}M_{A, iakm}M_{B, ibns}\ket{A, k, B, n, c}.
\label{generalresult}
\end{align}
%
%
The action of a simple idempotent $p_{C}^{cc}$ on such a state is therefore 
%
%
\beqn 
p_{C}^{cc}&\ket{A,a,B,b,c}=\sum\limits_{\substack{i,j, s, r,\\ k, m, n}} M^{-1}_{C, iccj}F^{ii0}_{aar}F^{ii0}_{bbs}F^{sib}_{air}F^{src}_{ijk}\nonumber \\ &\times F^{skj}_{cin} F^{iar}_{ikm}\sqrt{\frac{d_i d_a}{d_r}}M_{A, iakm}M_{B, ibns}\ket{A, k, B, n, c}\nonumber\\ 
 &= \sum\limits_{k, n} \mathcal{M}_{kn, ab}\ket{A, k, B, n, c},
 \label{eq:pCcc}
\eeqn
%
%
where, in the last line, we have introduced the matrix-like notation:
%
%
\begin{eqnarray}
\mathcal{M}_{kn, ab}(A,B,C,c)&=&\sum_{\substack{ i,j, s,\\ r, m}}F^{ii0}_{aar}F^{ii0}_{bbs}F^{sib}_{air}  F^{src}_{ijk}F^{skj}_{cin}F^{iar}_{ikm}\nonumber \\    
&\times& \sqrt{\frac{d_i d_a}{d_r}}M^{-1}_{C, iccj} M_{A, iakm}M_{B, ibns}.\nonumber 
\end{eqnarray}
%
%
An important property of Eq.~(\ref{eq:pCcc}) is that $A$, $B$ and $c$ are fixed and that only $a$ and $b$ are mixed into $k$ and $n$. 

The identity operator is obtained by summing Eq.~(\ref{eq:pCcc}) over all $c$ and $C$:
%
%
\be
\sum_{C, c}p_C^{cc}\ket{A,a,B,b, c}=\ket{A,a,B,b,c}.
\ee
%
%
We can construct the eigenvectors of $p_C^{cc}$ by solving the equation 
%
%
\be
\sum_{a, b}\mathcal{M}_{kn, ab} \, y_{ab} = y_{kn},
\ee
%
%
for all $y_{kn}$, where 
%
%
\be
\ket{A, B, C, c} = \sum_{a,b}y_{ab}\ket{A, a, B, b, c}.
\ee
%
%
Since for every quasiparticle $C$ there is a unique linear combination of the states $\ket{A, a, B, b, c}$, the trace of $p_{C}^{cc}$ over $a$ and $b$ is always $1$ (if $C$ is a fusion outcome of $A$ and $B$, i.e., the state $\ket{A, B, C, c}$ exists), and zero otherwise:
%
%
\be
\begin{split}
\sum_{a,b}\bra{A, a, B, b, c}p_C^{cc}\ket{A, a, B, b, c}= \sum_{a, b}\mathcal{M}_{ab, ab}\\ = \left\lbrace\begin{matrix}
\begin{array}{l}
1,\textnormal{ if } C\in A\times B\\
0,\textnormal{  otherwise}.
\end{array}
\end{matrix}\right.
\label{trace}
\end{split}
\ee
%
%
This is the key equation that allows one to obtain the fusion rules, i.e., $N_{A,B}^C$ of the Drinfeld center from the tube algebra.

We also have 
%
%
\be
\textnormal{Tr}_{a, b}(P_C) = \sum_{a, b, c}\mathcal{M}_{ab,ab} = n_C. 
\ee
%
%
The dimension of the two-quasiparticle Hilbert space can be recovered by computing
%
%
\be
\dim \mathcal{H}_{2\textnormal{QP}}=\sum_{A, B, C, a, b, c}\mathcal{M}_{ab, ab}.
\ee
%
%

%
%
\subsubsection{Fusion of a pure fluxon with a fluxon}
%
%

We now aim at showing that the product of a pure fluxon with another fluxon necessarily gives a fluxon. In the following, we will take $A$ to be a pure fluxon. All its tubes are contained only in the sector $11$, so that Eq.~\eqref{trace} simplifies to (using identities on the $F$-symbols):
%
%
\be
\delta_{c,b}\sum_{i, j}M^{-1}_{C, ibbj} \frac{1}{d_i}M_{A,i11i}M_{B,ibbj}=\left \lbrace\begin{matrix} 
\begin{array}{l}
1,\textnormal{ if } C\in A\times B\\
0, \textnormal{ otherwise}.
\end{array}
\end{matrix}\right.
\label{easy}
\ee 
%
%
for any sector $b$ of $B$. In particular, for $b=1$, we have
%
%
\be
\sum_{i}M^{-1}_{C, i11i} \frac{1}{d_i}M_{A,i11i}M_{B,i11i}=1
\label{supereasy}
\ee 
%
%
when $C\in A\times B$. If $B$ is a fluxon, there is necessarily some $i$ for which $M_{B, i11i}$ is nonvanishing. In order to satisfy equation Eq.~\eqref{supereasy}, we then see that $C$ must have some nonvanishing $M^{-1}_{C, i11i}$, which means it has weight on the $11$ sector and is also a fluxon.\\

We can go further when the pure fluxon $A$ is the vacuum. In this case, we have a simple expression for $M_{\mathbf{1}, i11i}=d_i$, so that Eq.~\eqref{easy} becomes 
%
%
\be
\sum_{i,j}M^{-1}_{C, ibbj}M_{B, ibbj}=1.
\label{fusionvac}
\ee
%
%
Using Eq.~\eqref{generalid}, we see that Eq.~\eqref{fusionvac} is true only when $C=B$, as it is expected for the fusion with the vacuum.

%
%
\subsubsection{Example: Fibonacci}
%
%
As an example, let us consider the case where $\mathcal{C}$ is the Fibonacci category. It is a non-Abelian UMTC which contains two objects, $1$ and $\tau$.
As a UMTC, its Drinfeld center is built as the direct product $\mathcal{C} \times \overline{\mathcal{C}}$ where $\overline{\mathcal{C}}$ is the mirror image of   $\mathcal{C}$ (opposite chirality) (see, e.g., Ref.~\cite{Rowell09} for more details). 

There are five one-quasiparticle states $|A,a\rangle$: $\ket{(1,1), 1}$, $\ket{(1,\tau), \tau}$, $\ket{(\tau, 1), \tau}$, $\ket{(\tau, \tau), 1}$, $\ket{(\tau, \tau), \tau}$, while there are $34$ two-quasiparticle states.
Let us  look in particular at the case where $A = (\tau, \tau)$, $B = (\tau, \tau)$ and $c = 1$. In this subspace, we have two states written in the $\ket{A, a, B, b, c}$ basis as: 
%
%
\be
\begin{split}
\ket{a_1} =\ket{(\tau, \tau), 1, (\tau, \tau), 1, 1}, \\
\ket{a_2} = \ket{(\tau, \tau), \tau, (\tau, \tau), \tau, 1},
\end{split}
\ee
%
%
while in the basis $\ket{A, B, C, c}$ the two states are
%
%
\be
\begin{split}
\ket{b_1} = \ket{(\tau, \tau), (\tau, \tau), (1, 1), 1}, \\
\ket{b_2} =\ket{(\tau, \tau), (\tau, \tau), (\tau, \tau), 1}.
\end{split}
\ee
%
%
The relation between the two bases are: 
%
%
\begin{align}
\ket{b_1} &= \frac{1}{\sqrt{2}}(\ket{a_1} - \ket{a_2}),\\
\ket{b_2} &= \sqrt{\frac{2}{5+\sqrt{5}}}\quad \ket{a_1} + \sqrt{\frac{5+\sqrt{5}}{10}}\quad \ket{a_2}.
\end{align}
%
%
For other choices of  $A$, $B$ and $c$ there might exist only one state in which case the two bases are equivalent. For example, when $A = (1,1)$, $B=(1,1)$ and $c=1$,one  has 
%
%
\be
\ket{(1,1), 1, (1,1), 1, 1}= \ket{(1,1), (1,1), (1,1), 1}.
\ee
%
%

%
%
\section{Surgery approach to degeneracies}
\label{app:surg}
%
%
In this appendix, our objective is to calculate the degeneracy of a state of a (2+1)-dimensional TQFT on a $g$-handled torus $\Sigma_g$, with the following possible complications:  (a) we may have quasiparticles at fixed positions on the torus; (b) we may have WWL operators running around one of the (space-like) handles.   We will do this calculation with some nice TQFT and topological techniques.    Our calculation will give the topological degeneracies of a pure TQFT and will not account for the nontopological degeneracies of the SN model.  Hence our result will only be precise for the case of the Drinfeld center of a commutative UFC where there are no nontopological degeneracies. For the quantum double of noncommutative UFCs, the nontopological degeneracies would have to be added to the expressions by hand. In this way, we will recover many results of the main text in an alternative way and without explicit reference to a specific lattice model such as the SN or RSN.

We will consider a (closed) three-dimensional space-time manifold of the form ${\cal M} = \Sigma_g \times S^1$, where $\Sigma_g$ is the $g$-handled torus which will be our space-like manifold and $S^1$ is a time-like circle.   We are going to describe this manifold using ``surgery."    It turns out (see Refs.~\cite{Gompf_book,Kirby_book}) that any orientable 3-manifold can be obtained by starting with the 3-sphere $S^3$ and doing {\it surgery} on some link (in the sense of knot theory) embedded in $S^3$ (this fact is known as the ``Lickorish-Wallace Theorem"\cite{Lickorish1,Lickorish2,Wallace}).    By ``surgery" we mean the following procedure (see the discussion in Refs.~\cite{Simon_book,Gompf_book} for example):  (a) Thicken each strand of the link into a solid torus $S^1 \times D^2$ with $D^2$ the disk  (b)  For each strand remove this solid torus from the manifold and  (c) for each $S^1 \times D^2$ removed, insert instead a new solid torus $D^2 \times S^1$, i.e., switch the contractible and noncontractible directions. 

\begin{figure} [h] 
\includegraphics[angle=270,width=.8\columnwidth]{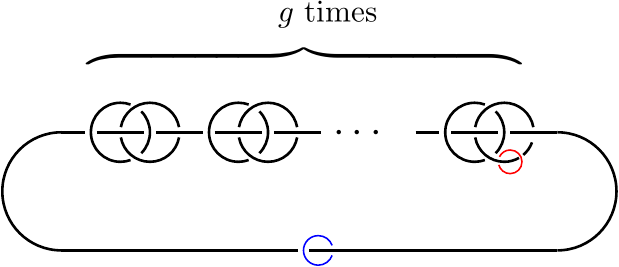}
\caption{Without the small blue and red loops, this link (0-framed, meaning no line has any self twists) is the surgery presentation of the 3-manifold $\Sigma_g \times S^1$ where $\Sigma_g$ is a $g$-handled torus.  The blue loop is a loop in the $S^1$ or time-like direction after surgery.  The red loop is a noncontractible space-like loop on the $g$-handled torus after surgery. }
    \label{fig:gknotfigure}
\end{figure}

It is a known fact from topology that the manifold $\Sigma_g \times S^1$ is given by surgery on the knot shown in Fig.~\ref{fig:gknotfigure} (only the black pieces) where all the strands are 0-framed, meaning they have no self-twists.  This fact is given explicitly (and the derivation explained) in Fig.~6.4 of Ref.~\cite{Gompf_book} (to be precise, in that figure they are using the same diagram to describe the 4-manifold that is bounded by our 3-manifold of interest, however the diagram is the same to just describe the 3-manifold, see also Ref.~\cite{Kirby_book}).  To give some intuition behind this topological statement, surgery on a single loop turns $S^3$ into the manifold $S^2 \times S^1$, i.e., it generates one noncontractible loop (the $S^1$) (see Ref.~\cite{Simon_book} chapter 24 for example).    Similarly, if we ignore the long string in Fig.~\ref{fig:gknotfigure} we have $2g$ unlinked loops, and surgery on all of these unlinked loops generates the so-called ``connected-sum" of $2 g$ factors of $S^2 \times S^1$. (A ``connected-sum" of two three-manifolds ${\cal M}_1$ and ${\cal M}_2$ is given by removing a ball from each manifold to form ${\cal M}_1 \char`\\ B^3$ and ${\cal M}_2\char`\\B^3$ with the notation $\char`\\$ meaning ``remove".  Then we sew ${\cal M}_1\char`\\B^3$ and ${\cal M}_2\char`\\B^3$  together on the spherical surfaces that have been exposed to form the connected-sum which is notated ${\cal M}_1 \# {\cal M}_2$ ).   The resulting manifold has $2g$ independent noncontractible loops.   The small red loop, after the surgery, goes around one of these noncontractible loops.   The final surgery on the long loop adds one more noncontractible direction (this is the time-like $S^1$) and after surgery the blue loop goes around this noncontractible direction.   Note that this last loop is now not independent of the other noncontractible loops --- and indeed, ties the other directions together in pairs --- and generates a nontrivial fundamental group after the surgery.   For the case of $g=1$, the knot in Fig.~\ref{fig:gknotfigure} is the Borromean rings and surgery gives the 3-torus with $\Sigma_g$ just the usual single-handled torus\cite{Gompf_book,Kirby_book}.

Now, as discussed in Refs.~\cite{Simon_book,Witten89} the degeneracy of states on $\Sigma_g$ is given by the partition function of the TQFT on the 3-manifold $\Sigma_g \times S^1$
\begin{eqnarray} \nonumber
 \mbox{Number of states }(\Sigma_g)  &=& {\rm Tr} \left[ \sum_\text{states} |{\mbox{state}} \rangle \langle  {\mbox{state}}  |  \right],  \\ &=& {\newZ}(\Sigma_g \times S^1).\label{eq:numberofstates}
\end{eqnarray}
Here we have used the notation ${\newZ}$ for a TQFT partition function, as for example used by Witten~\cite{Witten89}.  We use mathscript font here so as not to confuse this quantity with the finite-temperature partition functions $Z$ in the main text or the Drinfeld center $\cal Z$. 

The expression \eqref{eq:numberofstates}  remains true even if there are particles at fixed positions on $\Sigma_g$ (we just absorb these particles into the definition of $\Sigma_g$ and it is exactly the same equation).   For a WWL operator labeled with particle type $w$ going around a (space-like) noncontractible loop $L$, we need to add such a loop into the 3D manifold when we calculate the partition function
\begin{eqnarray} \label{eq:wilson1}
     \langle W_w(L)\rangle &=&   {\rm Tr} \left[ W_w(L) \sum_\text{states} |{\mbox{state}} \rangle \langle  {\mbox{state}}  |  \right], 
 \\ & =& {\newZ}(\Sigma_g \times S^1  \mbox{ 
   with loop $w$ around handle $L$}).  \nonumber
\end{eqnarray}

To calculate these partition functions we use the Reshetikhin-Turaev~\cite{Reshetikhin91,Simon_book} approach, that is, we represent the manifold $\Sigma_g \times S^1$ with the above-mentioned surgery presentation, we label each surgery loop with the Kirby strand $\Omega$  [see Eq.~\eqref{eq:OmegaDef}] and then simply calculate the diagram that results using the usual TQFT (modular tensor category) evaluation rules.   If there are also additional particle lines in the picture (say the red or blue lines going around handles) these can be evaluated using the TQFT evaluation rules as well.   The general result is
\be \label{eq:RTequation}
 {\newZ}({\cal M}, \mbox{labeled link}) = 
 \frac{
 e^{i \phi}}{\cal D}  \left(  \mbox{diagram evaluation}  \right), 
\ee
where ${\cal D} = \sqrt{\sum_a d_a^2}$ is the total quantum dimension of the TQFT and the phase $\phi$ for our purposes is going to be zero. In general the phase depends on the so-called ``signature" of the surgery $\Omega$ link.  However that is zero here since none of the strands are linked with each other and none of the strands have self-twists.  

\begin{figure}[h]
    \includegraphics[width=.6\columnwidth]{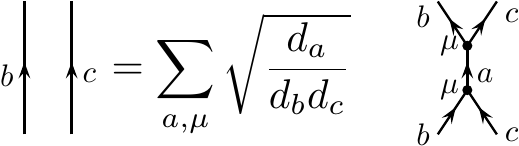}
  \caption{Completeness relation.}
  \label{fig:completeness}
\end{figure}

\begin{figure}[h]
    \includegraphics[width=.5\columnwidth]{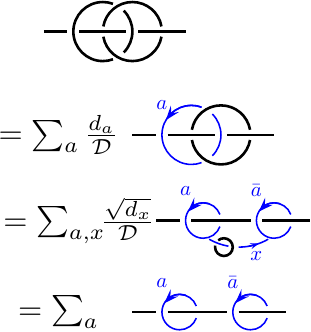}
\caption{The black lines are $\Omega$ strands.  Going from the first line to the second inserts the definition of $\Omega$ for one loop.  Going to the third line uses the completeness relation shown in Fig.~\ref{fig:completeness} (vertex indices are not written for simplicity).  Going to the final line uses the killing property.}
\label{fig:doublering}
\end{figure}

In evaluating the diagram we will make extensive use of the so-called ``killing" property of the $\Omega$
loop:  The value of an (untwisted) $\Omega$ loop is $\cal D$ if there is no net particle going through the loop (i.e., the total quantum number going through is zero) and the value of the diagram is zero otherwise.  Using the definition of $\Omega$, the completeness relation of Fig.~\ref{fig:completeness}, and the killing property we have the graphical equality shown in Fig.~\ref{fig:doublering}

\begin{figure}[h]
  \includegraphics[width=.6\columnwidth]{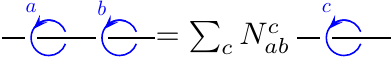}
\caption{Fusion of loops.  See Ref.~\cite{Simon_book}, Figs. 17.1 and 17.3.}
\label{fig:mergering}
\end{figure}

Thus  using the relation in Fig.~\ref{fig:doublering} a total of $g$ times, the picture in Fig.~\ref{fig:gknotfigure} (ignoring red and blue loops) is reduced to a very simple diagram with a single long $\Omega$ loop linked by $2g$ loops labeled $a_1, \bar a_1, a_2, \bar a_2, \ldots a_g, \bar a_g$ with a sum over all the $a$'s.   We can then merge the loops with each other using the identity shown in Fig.~\ref{fig:mergering}.

Once all of the $a$ loops are fused together, we use  Eq.~\eqref{eq:RTequation} to 
obtain the ground-state degeneracy of a $g$-handled torus: 
\begin{eqnarray}
\nonumber
 {\newZ}(\Sigma_g \times S^1) &=& \frac{1}{\cal D} \sum_{a_1, \ldots, a_g, b} N_{a_1 \bar a_1 a_2 \bar a_2, \ldots a_g \bar a_g}^{b}  \raisebox{-2mm}{\includegraphics[]{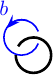}},
 \\
&=& \rule{0pt}{15pt}\sum_{a_1, \ldots, a_g} N_{a_1 \bar a_1 a_2 \bar a_2, \ldots a_g \bar a_g}^{\bf 1}. 
\label{eq:Nsum}
\end{eqnarray}
Note that the black $\Omega$ loop in Eq.~\eqref{eq:Nsum} is the long
$\Omega$ loop from Fig.~\ref{fig:gknotfigure}.  Going to the last line
in Eq.~\eqref{eq:Nsum} we have used the killing property again.  Note
that this results agrees with Eq.~\eqref{eq:multifusionN} in the absence
of plaquette fluxes.

We now consider how this calculation changes if we add quasiparticles at fixed positions.  Such quasiparticles remain in the same position at all time and thus wrap around the time-like $S^1$ direction of $\Sigma_g \times S^1$.  These correspond to the blue ring in Fig.~\ref{fig:gknotfigure}.   When we calculate ${\newZ}({\cal M}, \mbox{labeled link})$ the diagram to be evaluated now includes one additional labeled ring around the long $\Omega$ loop for every quasiparticle present.    These rings may be fused together using Fig.~\ref{fig:mergering} and then also fused with the $a, \bar a$ rings as well.

Thus if we have quasiparticles labeled $A_1, \ldots, A_M$ on our torus, the  degeneracy of states is now
\beqn
{\newZ}(\Sigma_g \times S^1; A_1, \ldots A_M) = \phantom{BIG SPACE}  \\  
\sum_{a_1, \ldots a_g} N_{a_1 \bar a_1 a_2 \bar a_2 \ldots a_g \bar a_g A_1, A_2, \ldots, A_M}^{\bf 1}, \nonumber
\eeqn
which agrees with Eq.~\eqref{eq:multifusionN}.


Now we consider a WWL operator around a space-like handle, say labeled with $w$ as we consider in Eq.~\eqref{eq:wilson1}.  This  corresponds to a labeled loop like the red loop in Fig.~\ref{fig:gknotfigure}.  Here, one of our $a$ loops that we obtain from Fig.~\ref{fig:mergering} will end up with a red $w$-loop around it.   However,  we can then use the unwrapping move shown in Fig.~\ref{WWloop} to remove the $w$ loop and accumulate a factor of $S_{w.a}/S_{{\bf 1},a}$.  Thus we obtain
\begin{align} \nonumber
&  {\newZ}(\Sigma_g \times S^1; A_1, \ldots, A_M; \mbox{\small loop $w$ around handle 1}) = \\
& ~~~~~~\sum_{a_1, \ldots a_g} \frac{S_{w a_1}}{S_{{\bf 1},a_1}} N_{a_1 \bar a_1 a_2 \bar a_2 \ldots a_g \bar a_g A_1, A_2, \ldots, A_M}^{\bf 1}.  \label{eq:onehandle}
\end{align}
To compare to the main text we convert the WWL operator to
obtain a projector $P_X$ using Eq.~\eqref{eq:PW}.  This then tells us
that the dimension of the space such that only flux $X$ goes around the handle 1 is given by
\beqn
 \langle P_X(L_1)\rangle = \sum_{a_2, \ldots a_g} N_{X \bar X a_2 \bar a_2 \ldots a_g \bar a_g A_1, A_2, \ldots, A_M}^{\bf 1}, 
\eeqn
which also matches the main text equations \eqref{eq:multifusionN} and \eqref{eq:increasegenus}.

Note that with this approach we can also calculate the expectation of multiple Wilson loops going around different handles.  For example, 
\begin{align}
&  {\newZ}(\Sigma_g \times S^1; A_1, \ldots, A_M; \mbox{\small loop $w_1,w_2$ around handles 1,2}) = \nonumber \\
& ~~~~~~\sum_{a_1, \ldots a_g} \frac{S_{w_1 a_1}} {S_{{\bf 1},a_1}}\frac{S_{w_2 a_2}}{S_{{\bf 1},a_2}} N_{a_1 \bar a_1 a_2 \bar a_2 \ldots a_g \bar a_g A_1, A_2, \ldots, A_M}^{\bf 1}.
\end{align}

It is interesting to consider a WWL that follows a throat as
in Fig.~\ref{fig:genus1}.  Let us first imagine that on one side of
the throat is a surface of genus 1 and the other side is genus $g-1$. 

\begin{figure}
    \centering
    \includegraphics{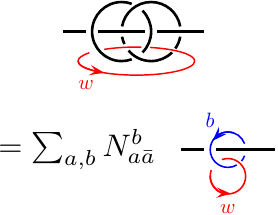}
    \caption{A WWL around a throat that separates a single
      handle from the remainder of the handles.}
    \label{fig:nullhomol}
\end{figure}

Let $p$ be a path that follows one nontrivial handle in
Fig.~\ref{fig:genus1} and let $q$ be the path following the conjugate
handle direction.  The path around the cut disk where we put $w$ can
be represented as $p q p^{-1} q^{-1}$.  In our surgery diagram we
would have a WWL as shown in the upper half of
Fig.~\ref{fig:nullhomol}.  Using the same techniques as in
Figs.~\ref{fig:doublering} and \ref{fig:mergering} we can evaluate
the diagram to obtain the expression in the lower half.  The $w$ loop
can be removed using the unlinking move to accumulate a factor of
$S_{w b}/S_{{\bf 1}, b}$ and we end up with the final result of
\begin{align}
&  {\newZ}(\Sigma_g \times S^1; A_1, \ldots, A_M; \mbox{\small loop $w$ around $pqp^{-1}q^{-1}$}) = \nonumber \\
& ~~~~~~\sum_{b, a_1, \ldots a_g} \frac{S_{w b}}{S_{{\bf 1}, b}} N_{a_1 \bar a_1}^b N^{\bf 1}_{b a_2 \bar a_2 \ldots a_g \bar a_g A_1, A_2, \ldots, A_M}.
\end{align}

If we instead have our WWL follow a throat that separates
genus $g_\mathcal{R}$ from genus $g_{\overline{\mathcal{R}}}$ on a $g=g_\mathcal{R} + g_{\overline{\mathcal{R}}}$ genus surface, then
WWL path can be represented as
\be
p_1 q_1 p_1^{-1} q_1^{-1} p_2 q_2 p_2^{-1} q_2^{-1} \ldots p_{g_\mathcal{R}} q_{g_\mathcal{R}} p_{g_\mathcal{R}}^{-1} q_{g_\mathcal{R}}^{-1},
\ee
where $p_j$ and $q_j$ are the two conjugate paths around the $j^{th}$
handle.  The link analogous to the top of Fig.~\ref{fig:nullhomol}
would have a single red strand threading through $g_\mathcal{R}$ pairs of
$\Omega$ loops before closing.  We use the same procedure to
evaluate the handle and obtain a figure similar to the lower of
Fig.~\ref{fig:nullhomol} except now the $w$ strand is linked to $g_\mathcal{R}$
different $b$ strands which we call $b_1, \ldots, b_{g_\mathcal{R}}$.  The $b$
strands can be fused together to one single loop we call $b$, then the
unlinking move can be invoked.  The result of this calculation is
\begin{align}
  &  {\newZ}(\Sigma_g \times S^1; A_1, \ldots, A_M;\nonumber \\
  &~~~~~~~\mbox{\small loop $w$ around throat separating $g_\mathcal{R}$ from $g_{\overline{\mathcal{R}}}$}) = \nonumber\\
  & ~~~\sum_{b, a_1, \ldots a_g} \frac{S_{w b}}{S_{{\bf 1}, b}} N_{a_1 \bar a_1 a_2 \bar a_2 \ldots a_g \bar a_{g_\mathcal{R}}}^b   \\
   & ~~~~~~~~~~~~~~~~~~~~~~~~~~~~~ N^{\bf 1}_{b a_{g_\mathcal{R} + 1}  \bar a_{g_\mathcal{R} + 1}  \ldots a_g \bar a_g A_1, A_2, \ldots, A_M}, \nonumber
\end{align}
Converting this result from a WWL to a projector $P_X$ using Eq.~\eqref{eq:PW} we obtain
\be
N_{a_1 \bar a_1 a_2 \bar a_2 \ldots a_g \bar a_{g_\mathcal{R}}}^X   N^{\bf 1}_{X a_{g_\mathcal{R} + 1}  \bar a_{g_\mathcal{R} + 1}  \ldots a_g \bar a_g A_1, A_2, \ldots, A_M},
\ee
which matches Eqs.~\eqref{eq:multifusionN}. and \eqref{eq:dimXdimX} except that all of the
plaquette fluxes have ended up on the $g_{\overline{\mathcal{R}}}$ side. The reason for this
is that we arranged for our WWL link to surround the handles
(by linking the red line through the relevant $\Omega$ strands), but
we did not arrange for them to surround the plaquette fluxes.  To move
plaquette fluxes to the other side we simply need to link the red
WWL through the plaquette flux loops (blue in
Fig.~\ref{fig:gknotfigure}) as well.

A more interesting case to consider is where we have two Wilson loops going around ``conjugate" handles, i.e., where projected to the two-dimensional surface, the two loops would have to intersect.  
First, we realize that the temporal order of the two loops cannot matter, since the time-like direction is periodic.   We then want to consider Wilson loops around {\it both} rings at the top of Fig.~\ref{fig:doublering} to indicate that they go around the two conjugate handles created by these two $\Omega$ rings after surgery.  Let us consider putting a $w_1$ loop around the left loop at the top of Fig.~\ref{fig:doublering} and a $w_2$ loop around the right ring.    
\begin{figure}[t!]
\includegraphics[]{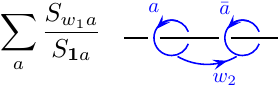}
   \caption{With Wilson loops $w_1$ and $w_2$ around two conjugate handles, instead of Fig.~\ref{fig:doublering} one obtains this. }
    \label{fig:altdoublering}
\end{figure}
We follow the same procedure to the second line of Fig.~\ref{fig:doublering} and we can again remove the $w_1$ ring around the $a$ loop with an unlinking factor of $S_{w_1 a}/S_{{\bf 1} a}$.  Going to the third line of Fig.~\ref{fig:doublering} the $w_2$ loop links the small $\Omega$ loop.   Fortunately, we can use Fig.~\ref{fig:completeness} to fuse $w_2$ with $x$ inside the small $\Omega$ loop and then use the killing property of $\Omega$ to obtain Fig.~\ref{fig:altdoublering} instead of the final diagram of Fig.~\ref{fig:doublering}.  Further evaluation is fairly straightforward : fuse all other $a$'s into a single quantum number $b$ and then one has a single (long) $\Omega$ loop with three loops wrapped around it:  $b$ as well as the $a$ and $\bar a$ from Fig.~\ref{fig:altdoublering}.  These three can be fused together using Fig.~\ref{fig:completeness} and one can use the killing property of the last $\Omega$.  The diagram that remains is a single $F$ symbol. 

A simple case to consider is the case of a 1-handle torus with WWL  $w_1$ and $w_2$ around the two conjugate noncontractible loops.  The simplest way to get this is to realize that $\Sigma_g \times S^2$ is a 3-torus, and we can put WWL  $w_1$ and $w_2$ around any two of the noncontractible loops --- so let us put one around a spatial loop and one around a time-like loop.   Thus, we can invoke the result of Eq.~\eqref{eq:onehandle} to give us:
\beqn
  \sum_{a}  \frac{S_{w_1 a}}{S_{{\bf 1} a}}  N_{a \bar a w_2} .
\eeqn
This does not look very symmetric between the two loops.  However, using the Verlinde form of $N$ we obtain:
\beqn
 \sum_{a,b} \frac{S_{w_1 a} S_{w_2 b} S_{a b} S_{\bar a b}}{S_{{\bf 1} a} S_{{\bf 1} b}} .
\eeqn

%
%
 \section{Relation to the 1D  classical Potts model}
\label{app:potts}
%
%
The $q$-state  classical Potts model has the following partition function:
\be
Z_\text{Potts}=\lambda_q^{N_\text{p}}+(q-1)\lambda_0^{N_\text{p}},  \text{ with } \lambda_q = {\rm e}^\beta + q-1.
\ee
In the thermodynamic limit, it becomes the partition function of independent spins in a magnetic field:
\be
Z_\text{Potts}^\infty \simeq \lambda_q^{N_\text{p}}.
\ee
Corrections to the dominant term
\be
Z_\text{Potts}=Z_\text{Potts}^\infty \left[1+(q-1){\rm e}^{-N_\text{p}/N_\text{p}^*} \right]
\ee
 reveal a correlation length
 \be
 N_\text{p}^* = \frac{1}{\ln \frac{\lambda_q}{\lambda_0}}\simeq \frac{{\rm e}^\beta}{q} \to \infty,
 \ee
which diverges in the low-temperature limit. The fact that the correlation length is related to the ratio of the two largest eigenvalues of the transfer matrix is well-known in the context of spin chains~\cite{Kardar_book}.

For the SN model, the thermodynamic limit of the partition function $Z^\infty$ [given in Eq.~\eqref{eq:ZSNthermo}] is dominated by pure fluxons that have $q_C = \mathcal{D}\, n_{C,1}/d_C=\mathcal{D}$. The first correction is related to non-pure fluxons $C$ with the largest $q_C<\mathcal{D}$ (we call $\mathcal{N}$ this set).  Note that in some cases, (e.g., when $\mathcal{C}={\rm Vec}(G)$ for a finite group $G$), one may have $\mathcal{P}=\mathcal{F}$ and therefore $q_C=0$. Also, the set $\mathcal{N}$ may only contain one type of anyon. When $N_\text{p}\to \infty$, the SN partition function reads
\be
Z\simeq Z^\infty\left[1+\frac{\sum_{C\in \mathcal{N}} S_{\mathbf{1},C}^{2-2g}}{\sum_{A \in \mathcal{P}} S_{\mathbf{1},A}^{2-2g} }{\rm e}^{-N_\text{p}/N_\text{p}^*} +... \right],
\ee
which defines a correlation area
 \be
 N_\text{p}^* = \frac{1}{\ln \frac{{\rm e}^{\beta}+\mathcal{D}-1}{{\rm e}^{\beta}+q_C-1}}\simeq \frac{{\rm e}^\beta}{\mathcal{D}-q_C} \to \infty,
 \label{eqNpstaragain}
 \ee
that diverges in the  low-temperature limit. This is similar to the 1D Potts model with $q=\mathcal{D}$ except for $q_C$ which is not necessarily $0$. 
In the particular case where the input category is $\mathcal{C}={\rm Vec}(\mathbb{Z}_N)$, the SN partition function is
\be
Z=N^{2g-1} \, Z_\text{Potts},  \text{ with } q=N,
\ee
which makes the relation to the Potts model explicit. The above correlation area  \eqref{eqNpstaragain} is the same characteristic scale as the one appearing in the WWL [see Eq.~\eqref{eq:64}] and in the topological mutual information [see Eq.~\eqref{eq:Npstar}].

The divergence of the correlation area indicates a $T_c=0^+$ transition between a phase with topological order (at $T=0$) and a phase without (at $T>0$). The mechanism by which thermal fluctuations destroy topological order in the SN model is the same as in the 1D Ising (or Potts) model and related to a proliferation of point-like topological defects (fluxons in the SN model, domain walls in the Ising model), see for example~\cite{Dennis02,Bacon06,Nussinov08,Landon13}. The absence of interaction between fluxons (i.e., the fact that they are totally deconfined at $T=0$)  implies that there is no macroscopic energy barrier protection.


%

\end{document}